%% file: AAArticle.tex
\PassOptionsToPackage{
  a4paper,
  left=1.5cm,
  right=1.5cm,
  top=1.6cm,
  bottom=1.8cm
}{geometry}

\documentclass[pdflatex,sn-mathphys,Numbered]{sn-jnl}

\usepackage{graphicx}%
\usepackage{subcaption}
\usepackage{multirow}%
\usepackage{amsmath,amssymb,amsfonts}%
\usepackage{amsthm}%
\usepackage{mathrsfs}%
\usepackage[title]{appendix}%
\usepackage{xcolor}%
\usepackage{textcomp}%
\usepackage{manyfoot}%
\usepackage{booktabs}%
\usepackage{algorithm}%
\usepackage{algorithmicx}%
\usepackage{algpseudocode}%
\usepackage{listings}%
\usepackage{comment}
\usepackage{gensymb}
\usepackage{caption}
\usepackage{siunitx}
\DeclareSIUnit{\wtpercent}{wt.\%}
\DeclareSIUnit{\volpercent}{vol.\%}
\DeclareSIUnit{\atpercent}{at.\%}
\DeclareSIUnit{\HVhalf}{HV0.5}
\DeclareSIUnit{\rpm}{rpm}
\usepackage{hyperref}
\usepackage{booktabs}

\theoremstyle{thmstyleone}%
\theoremstyle{thmstyletwo}%

\theoremstyle{thmstylethree}%

\begin{document}

\title[Article Title]{Evolution of microstructure and crystallographic texture in a novel Zr--4Nb--2Al alloy during hot rotary swaging}

\author*[1]{\fnm{Adam} \sur{Strnad}}\email{adam.strnad@matfyz.cuni.cz}

\author[1]{\fnm{Jana} \sur{Šmilauerová}}

\author[1]{\fnm{Josef} \sur{Stráský}}

\author[2]{\fnm{Lukáš} \sur{Horák}}

\author[1]{\fnm{Jan} \sur{Čapek}}

\author[3]{\fnm{Alžběta} \sur{Endrychová}}

\author[1]{\fnm{Martin} \sur{Holčák}}

\author[3]{\fnm{Jiří} \sur{Zýka}}

\author[1]{\fnm{Miloš} \sur{Janeček}}

\author[1]{\fnm{Petr} \sur{Harcuba}}

\affil*[1]{\orgdiv{Department of Physics of Materials}, \orgname{Faculty of Mathematics and Physics, Charles University}, \orgaddress{\street{Ke Karlovu 5}, \city{Prague}, \country{Czech Republic}}}

\affil[2]{\orgdiv{Department of Condensed Matter Physics}, \orgname{Faculty of Mathematics and Physics, Charles University}, \orgaddress{\street{Ke Karlovu 5},\postcode{121 16}, \city{Prague}, \country{Czech Republic}}}

\affil[3]{\orgdiv{UJP PRAHA a.s.}, \orgaddress{\street{Nad Kamínkou 1345},\postcode{156 00}, \city{Prague}, \country{Czech Republic}}}


\abstract{A multi-step induction-assisted hot rotary swaging route was developed for the novel dual-phase Zr--4Nb--2Al alloy to investigate the evolution of microstructure, crystallographic texture, and mechanical properties during sequential deformation. Phase-transformation temperatures determined by electrical resistivity measurements were used to select processing conditions in the single-phase $\beta$ field at \SI{880}{\degreeCelsius} and the $\alpha+\beta$ field at \SI{780}{\degreeCelsius}. Induction heating with thermographic temperature control enabled reduction of the initial \SI{12}{\milli\metre} rod to \SI{6.3}{\milli\metre}, corresponding to a total area reduction of 72\%. Swaging in the $\beta$ field refined millimetre-sized prior-$\beta$ grains to approximately \SI{100}{\micro\metre}, while subsequent cooling produced a fine lamellar $\alpha+\beta$ microstructure. Further deformation in the $\alpha+\beta$ field disrupted the lamellar structure, promoted globularisation and recrystallisation of the $\alpha$ phase, and refined the $\beta$ phase. The microstructural evolution was accompanied by a pronounced $\langle 11\bar{2}0\rangle_{\alpha}\parallel\mathrm{SD}$ fibre texture in the $\alpha$ phase, whereas the $\beta$ phase exhibited predominantly $\langle 110\rangle_{\beta}\parallel\mathrm{SD}$ together with a $\langle 111\rangle_{\beta}\parallel\mathrm{SD}$ component. The phase constitution remained comparatively stable, with approximately 12--15~wt.\% retained $\beta$ throughout processing. Subsequent $\alpha+\beta$ swaging decreased the compressive yield strength from 1103 to \SI{839}{\mega\pascal} while increasing plastic deformability, with a maximum fracture strain of 32~\%. Radial hardness heterogeneity was progressively eliminated. These results demonstrate that induction-assisted rotary swaging in selected phase fields provides an effective route for tailoring microstructure, texture, and mechanical response in dual-phase zirconium alloys with potential nuclear applications.
}




\maketitle

\section{Introduction}\label{sec1}
Zirconium alloys are widely used as structural materials in nuclear applications due to their unique combination of low neutron absorption cross-section, adequate mechanical strength, and good corrosion resistance \cite{Garzarolli1996-in,motta_2015_corrosion}.
Conventional reactor-grade zirconium alloys, such as Zircaloy-type and Nb-containing alloys, are typically designed as predominantly $\alpha$-phase materials with only minor alloying additions (generally below \qty{3}{\wtpercent}), resulting in microstructures consisting of an $\alpha$-Zr matrix with fine precipitates or a small fraction of retained $\beta$ phase \cite{duan_current_2017}.

In contrast to conventional zirconium alloys, the development of duplex $\alpha+\beta$ microstructures has been increasingly explored as a means of achieving improved mechanical properties. Such microstructures enable significant strengthening compared to conventional $\alpha$-based alloys, with ultimate tensile strengths exceeding \qty{700}{\mega\pascal} reported in Zr--Sn--Nb systems \cite{veverkova_novel_2021}. This improvement in strength of dual-phase alloys can be attributed to phase-boundary strengthening \cite{yang_achieving_2026}, residual stress formation between the two phases \cite{Cai2012ZrNb}, load partitioning between the $\beta$ and $\alpha$ phases \cite{Cai2012ZrNb,Cai2009Compression}, and the refinement of $\alpha$ phase morphology during thermomechanical processing \cite{veverkova_novel_2021}. Moreover, it has been shown that phase boundaries can act as dislocation sources, improving the ductility and strain-hardening ability of alloys \cite{GAO2025116837,CUI2025112547,GAO2026285}, contributing to the favourable mechanical properties of fine-grained dual-phase materials.

The potential of tailoring dual-phase microstructures through thermomechanical processing is well established in the widely used titanium alloy Ti--6Al--4V, where processing in the $\beta$ and $\alpha+\beta$ phase fields enables precise adjustment of phase fraction and morphology, leading to optimised combinations of strength, ductility, and fatigue resistance \cite{weiss_modification_1986,elshaer_influence_2023}. Similarly, achieving microstructural refinement and enhanced mechanical properties in dual-phase $\alpha+\beta$ zirconium alloys requires careful optimisation of the thermomechanical processing sequence.

The mechanical behaviour of zirconium alloys is highly sensitive to microstructural evolution during processing, including phase transformations, recrystallisation, and texture development. In particular, in Zr--Nb alloys, the interaction between $\alpha$ and $\beta$ phases during thermomechanical processing plays a critical role in determining the resulting texture and deformation mechanisms ultimately affecting macroscopic mechanical properties \cite{Cai2012ZrNb}. These observations indicate that the traditional $\alpha$-based alloy design approach inherently limits the achievable mechanical properties of zirconium alloys, motivating the exploration of highly alloyed $\alpha+\beta$ zirconium alloys \cite{Kondo2011-si,Liu2022-ne}. This, in turn, motivates the development of thermomechanical processing strategies specifically tailored to such alloys. In particular, processing at elevated temperatures, guided by phase-transformation data, offers a route to control phase fraction, phase morphology, and the crystallographic texture of both the $\alpha$ and $\beta$ phases \cite{daniel_detailed_2019,Saxena2018-ho}.


Rotary swaging is a high-strain, incremental deformation technique widely used for the production of axisymmetric components, including rods and tubes. Its applicability to zirconium alloys used in nuclear fuel cladding has also been investigated \cite{Singh2022-ce}. The process enables substantial plastic deformation at relatively low forming forces, promoting pronounced grain refinement.

A key advantage of this technique fully utilized in this study is its compatibility with induction heating, which enables precise temperature control during deformation. This allows processing within selected $\beta$ and $\alpha+\beta$ phase fields, making rotary swaging a promising method for tailoring microstructure in multiphase alloys.

The effects of cold rotary swaging on $\alpha$-Zr alloys are well understood. In Zircaloy-4, cold swaging has been shown to produce significant grain refinement and improvements in both yield strength and ultimate tensile strength \cite{Singh2022-ce}. However, studies addressing rotary swaging of $\alpha+\beta$ zirconium alloys at elevated temperatures remain limited.

Only a few works have investigated warm or hot rotary swaging of zirconium alloys. Veverková et al. \cite{veverkova_novel_2021} applied rotary swaging to a novel $\alpha+\beta$ zirconium alloy preheated to \qty{900}{\degreeCelsius}, while Rogachev et al. \cite{rogachev_rotary_2022,rogachev_effect_2024} performed swaging at intermediate temperatures of \qty{200}{\degreeCelsius} and \qty{300}{\degreeCelsius} on Zr--2.5Nb, achieving an ultrafine-grained microstructure. Despite these contributions, there is currently no systematic study focused on rotary swaging of zirconium alloys purposefully processed at temperatures within the $\beta$ and $\alpha+\beta$ phase fields.

Due to its axisymmetric deformation mode combined with dominant axial material flow, rotary swaging promotes the development of a pronounced crystallographic fibre texture. In pure hcp materials with $c/a$ less than the ideal value, cold deformation results in a $\langle 10\bar{1}0\rangle$ fibre texture in the direction of deformation. This was shown for zirconium wire drawing \cite{Burgers1937_ZrWireTexture}, tube extrusion \cite{Nerses1960_ZrTubeTexture} and rolling~\cite{Liu2024-tb}. Cold rotary swaging was also shown to produce $\langle 10\bar{1}0\rangle$ fibre texture parallel to the swaging direction (SD) in Zr hcp $\alpha$ phase. This texture develops because the $\langle 10\bar{1}0\rangle$ directions tend to align with the deformation axis during rotary swaging, primarily as a consequence of prismatic $\{10\bar{1}0\}\langle 11\bar{2}0\rangle$ slip activation and subsequent grain reorientation, as reported for Zr~\cite{FU2023144584} and Ti~\cite{Meng2021-al,Hu2024-vt,Wang2018-fs}.

The evolution of texture of hcp Zr after cold deformation followed by recrystallisation annealing was studied in \cite{Chaubet2002_EBSDRecrystallizationZircaloy4,Tenckhoff1970_AnnealingTexturesZircaloy,Cheadle1966_TextureZrRichAlloys}. It was shown that, similarly to other hcp alloys with lower than ideal $c/a$ ratio \cite{Dillamore1965_PreferredOrientation}, the preferred orientation changes to $\langle 11\bar{2}0\rangle \parallel \mathrm{SD}$ after recrystallisation. However, recent study on cold rotary swaged commercially pure Zr showed that partial recrystallisation did not lead to the formation of this common texture, instead the $\langle 10\bar{1}0\rangle \parallel \mathrm{SD}$ present after the cold swaging was weakened as a result of recrystallisation annealing~\cite{FU2023144584}.






In pure bcc materials, the typical texture after cold rotary swaging is the $\langle 110\rangle \parallel \mathrm{SD}$ fibre texture which results from the $\{hkl\}\langle 111\rangle$ type slip being the dominant deformation mechanism. Formation of this texture in bcc tungsten was reported by Zhou \cite{Zhou2021-fo} and in $\beta$-type Ti alloy by Cheng \cite{Cheng2020-ut}.


As noted above, research on rotary swaging of zirconium alloys has predominantly focused on $\alpha$-based systems such as Zircaloy-4, while studies on dual $\alpha+\beta$ alloys (e.g., E125 or ternary Zr--Nb--Sn systems) remain limited. In particular, a systematic understanding of how controlled hot rotary swaging within the $\beta$ and $\alpha+\beta$ phase fields affects microstructural evolution and crystallographic texture is currently lacking.

The aim of this study was to develop a feasible hot rotary swaging route involving sequential deformation in the $\beta$ and in $\alpha+\beta$ fields and to investigate the evolution of crystallographic texture in a novel, highly alloyed dual-phase (hcp + bcc) zirconium alloy Zr--4Nb--2Al (\unit{\wtpercent}). The processing parameters were defined based on experimentally determined phase transformation data to ensure deformation within the targeted phase fields. Rotary swaging was carried out using a newly developed setup equipped with induction heating and thermographic monitoring, enabling precise control of temperature conditions throughout deformation.

\section{Material and experimental methods}\label{sec2}
The material investigated in this study is a novel dual-phase $\alpha+\beta$ Zr--4Nb--2Al alloy (\unit{\wtpercent}). In Zr alloys, niobium acts as a $\beta$-phase stabiliser \cite{Betterton1958-ci}, increases strength by solid-solution strengthening, and improves corrosion resistance while reducing hydrogen pickup \cite{Bell2017-wn}. These effects are particularly important for nuclear applications. In contrast, aluminium stabilises the $\alpha$ phase \cite{Jaworska2021-gd}, strengthens it through solid-solution hardening \cite{Zhang2020-cl}, and improves high-temperature performance by reducing creep \cite{MAGNUSSEN2026121891}. However, the aluminium content must remain limited to avoid the formation of ordered intermetallic phases such as Zr$_2$Al and Zr$_3$Al, which have been shown to adversely affect high-temperature oxidation resistance \cite{cox1967oxidation}.

\begin{figure}[h!]
    \centering
    \includegraphics[width=0.45\linewidth]{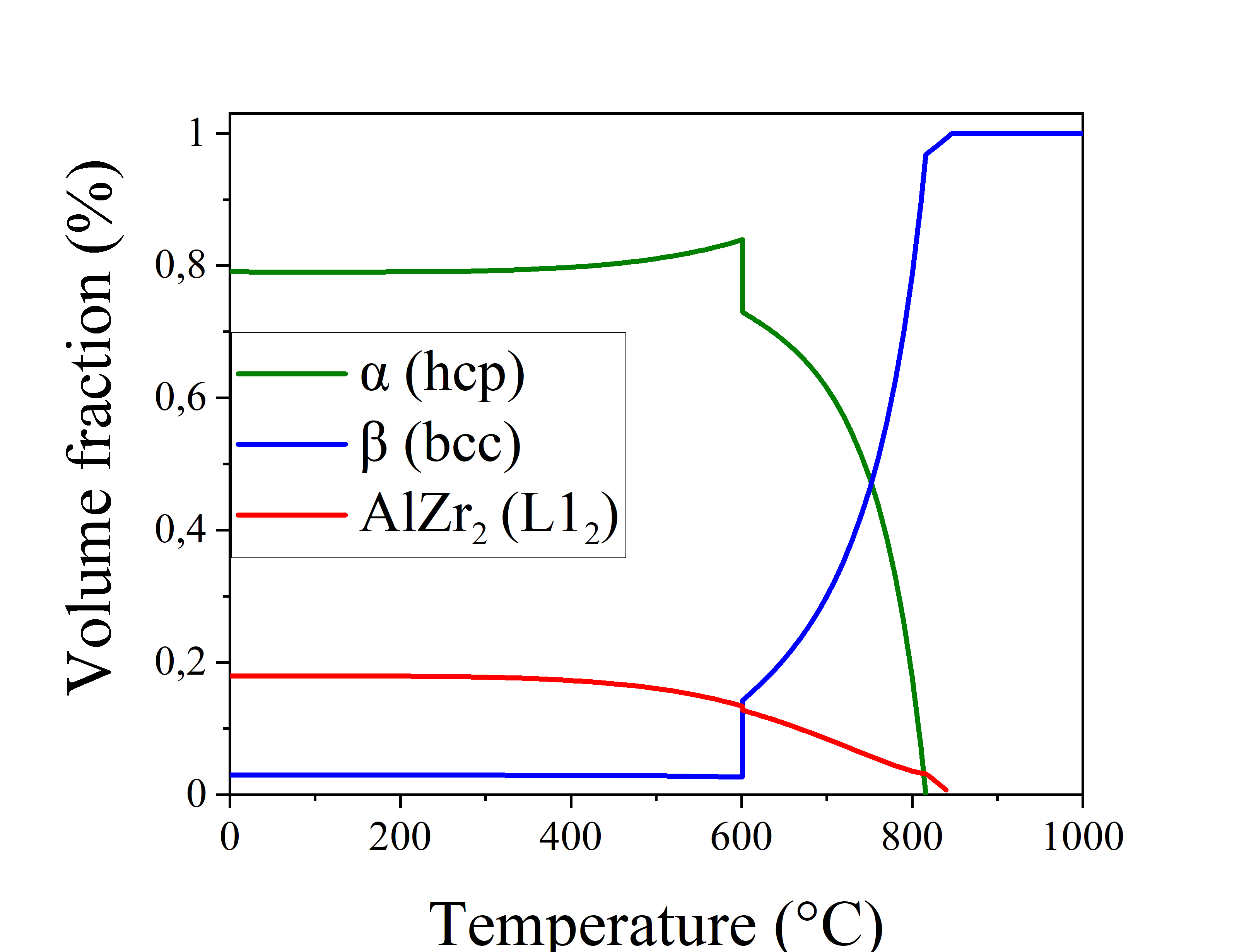}
    \caption{Theoretical equilibrium phase composition as a function of temperature calculated using Thermo-Calc software}
    \label{Thermo calc figure}
\end{figure}

This alloying concept leads to a thermodynamically stable $\alpha+\beta$ microstructure with a relatively high fraction of the $\beta$ phase compared with commercial alloys. Thermo-Calc simulations using High entropy library TCHEA8 indicate that, under equilibrium conditions, the alloy contains approximately \qty{5}{\volpercent} $\beta$ phase at room temperature. This fraction remains nearly constant up to \qty{600}{\degreeCelsius}, above which a marked increase in $\beta$-phase content is predicted. The $\beta$-transus temperature is estimated to be \qty{850}{\degreeCelsius} (see Figure~\ref{Thermo calc figure}). The simulated phase transformation data from Thermo-Calc were experimentally verified using resistivity measurement, which will be discussed later.

The Zr--4Nb--2Al alloy was produced at UJP Praha by electric arc melting and subsequently cast into a rod of \qty{12}{\milli\meter} diameter using a mono-arc furnace. Homogenisation annealing was carried out at \qty{1200}{\degreeCelsius} for \qty{4}{\hour} in an argon atmosphere, followed by slow furnace cooling. The material was then processed by hot rotary swaging.

\subsection{Hot rotary swaging}
\label{section: rod swaging}
\begin{figure}[!ht]
  \centering
  \includegraphics[width = 0.95\linewidth]{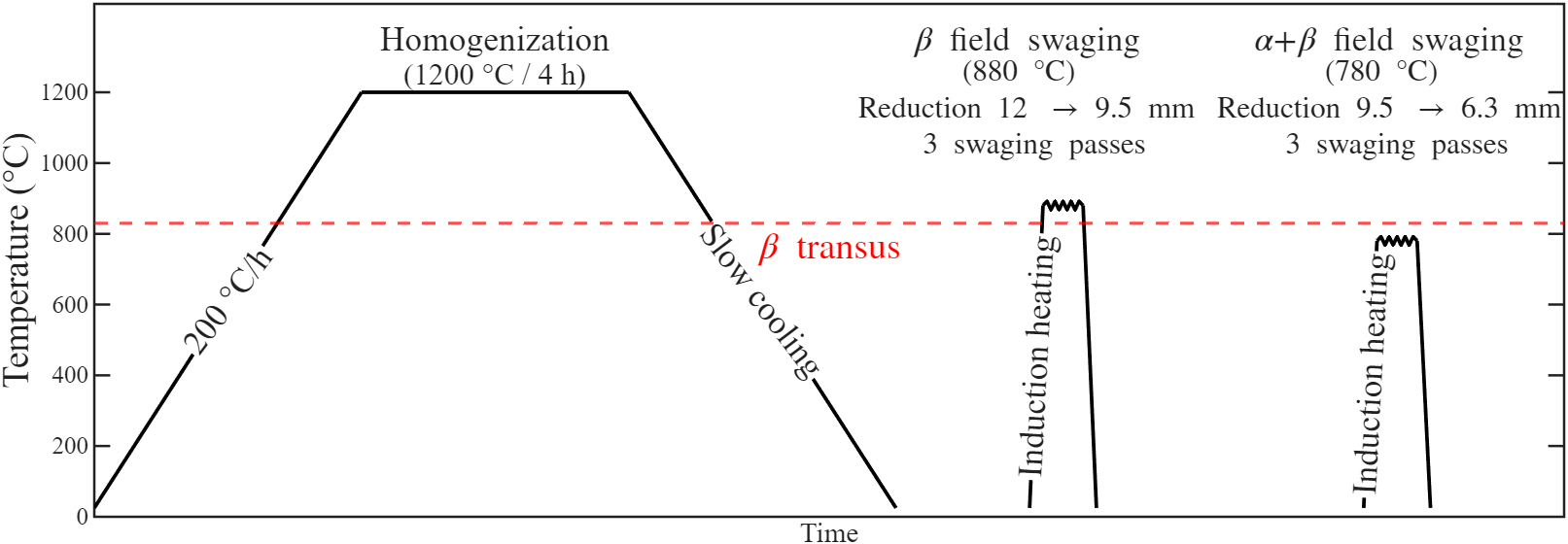}
  \centering
  \caption{Diagram of the rotary swaging process, consisting of homogenisation at \qty{1200}{\degreeCelsius} after casting, swaging in the $\beta$ field at \qty{880}{\degreeCelsius} followed by swaging in the $\alpha+\beta$ field.}
  \label{Swaging process graph}
\end{figure}

\begin{figure}[!ht]
    \centering

    \begin{subfigure}[t]{0.66\linewidth}
        \centering
        \begin{minipage}[c][4.2cm][c]{\linewidth}
            \centering
            \includegraphics[width=0.8\linewidth]{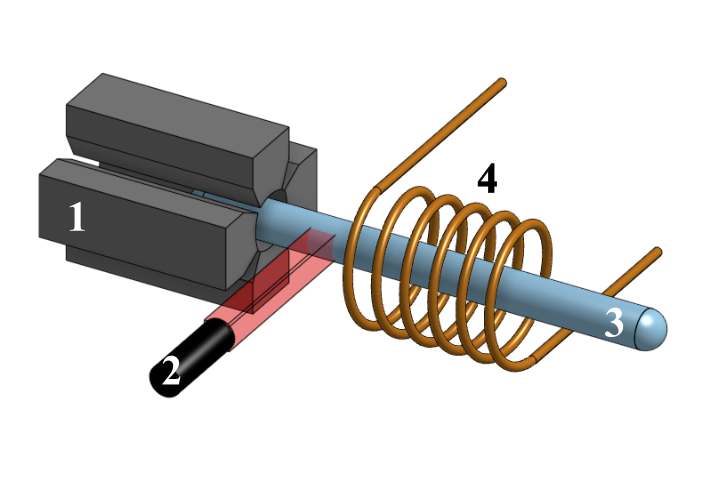}
        \end{minipage}
        \caption{ }
    \end{subfigure}
    \hfill
    \begin{subfigure}[t]{0.33\linewidth}
        \centering
        \begin{minipage}[c][4.2cm][c]{\linewidth}
            \centering
            \includegraphics[width=0.8\linewidth]{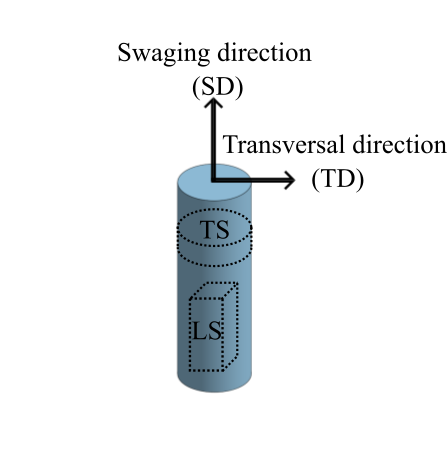}
        \end{minipage}
        \caption{ }
    \end{subfigure}

    \caption{Overview of rotary swaging. \textbf{a)} Diagram of swaging setup. 1) The swaging machine. 2) Thermo-camera used for temperature control. 3) Zirconium rod. 4) Induction coil for workpiece heating.  \textbf{b)} Coordinate system convention used. \textbf{TS}-transverse sample. \textbf{LS}-longitudinal sample.}
    \label{fig: swaging setup diagram}
\end{figure}
The complete processing route of the investigated alloy is shown in Figure~\ref{Swaging process graph}. The swaging was performed on an arc-melted and homogenised rod, which was then subjected to rotary swaging at predefined temperatures. The process was divided into two stages according to the target temperature: deformation in the single-phase $\beta$ region and deformation in the $\alpha+\beta$ dual-phase region.

The first stage, conducted in the $\beta$ region, aimed to refine the coarse millimetre-sized $\beta$ grains, present in the material after homogenisation. Such large grains would otherwise result in non-uniform mechanical properties along the rod. Moreover, due to the crystallographic orientation relationship governing the $\beta \rightarrow \alpha$ transformation ($\{110\}_\mathrm{bcc} \parallel  \{0001\}_\mathrm{hcp}$), the presence of large prior-$\beta$ grains would lead to a heterogeneous crystallographic texture in the transformed $\alpha$ phase \cite{Burgers1934}. 

The second stage was performed in the $\alpha+\beta$ coexistence region, where deformation promotes the development of $\alpha+\beta$ microstructure, dynamic recrystallisation, refinement and globularisation of the $\alpha$ phase \cite{Semiatin1999}.

The initial condition for rotary swaging was an arc-melted and homogenised rod with a diameter of \qty{12}{\milli\meter}, which was progressively reduced to a diameter of \qty{6.3}{\milli\meter} by rotary swaging. The process was carried out in discrete steps, 3 swaging passes in the $\beta$ field followed by 3 swaging passes in the $\alpha+\beta$ field, each pass reducing the diameter by approximately \qty{1}{\milli\meter}.

During deformation in the $\beta$ phase region, the diameter was reduced from \qty{12}{\milli\meter} to \qty{9.5}{\milli\meter}. Subsequent swaging in the $\alpha+\beta$ region further reduced the diameter from \qty{9.5}{\milli\meter} to \qty{6.3}{\milli\meter}. To investigate the evolution of microstructure and crystallographic texture, samples were extracted after reductions to diameters of \qty{9.5}{\milli\meter}, swaged only in the $\beta$ phase field, and \qty{8.3}{\milli\meter}, \qty{7.3}{\milli\meter}, and \qty{6.3}{\milli\meter}, swaged in the $\beta$ field followed by $\alpha+\beta$ swaging. These samples correspond to area reductions of \qty{37}{\percent}, \qty{52}{\percent}, \qty{63}{\percent}, and \qty{72}{\percent}, respectively. For brevity, the diameters are labelled as \qty{9}{\milli\meter}, \qty{8}{\milli\meter}, \qty{7}{\milli\meter}, and \qty{6}{\milli\meter} in the rest of this article. 

A schematic of the experimental setup is presented in Figure~\ref{fig: swaging setup diagram}a, while the coordinate naming convention used for further experiments on the swaged rods is displayed in Figure~\ref{fig: swaging setup diagram}b. The setup consisted of a HPS rotary swaging setup (1), a ThermoIMAGER TIM M1 Micro-Epsilon thermo-camera (2) for temperature monitoring, and an industrial induction heating unit (4). The processing temperature was controlled by manually adjusting the power of the induction heater based on feedback from the thermo-camera. Swaging was conducted at a constant feed rate of \qty{0.38}{\centi\meter\per\second}. To ensure reliable temperature measurements and a uniform surface emissivity, the rod was polished prior to each swaging step using a nonwoven abrasive. The emissivity of the prepared surface was determined to be 0.85 at \qty{800}{\degreeCelsius} and 0.80 at \qty{1000}{\degreeCelsius}.

These emissivity values were established by calibration against a thermocouple used as an independent temperature reference. For each calibration temperature, the polished sample was placed in a furnace preheated to either \qty{800}{\degreeCelsius} or \qty{1000}{\degreeCelsius}. Once thermal equilibrium had been reached, the furnace was opened and the emissivity setting of the thermal camera was adjusted until the indicated temperature matched that measured by the thermocouple.

\subsection{Experimental methods}
Microstructural characterisation was conducted using an Apreo 2 (Thermo Fisher Scientific) scanning electron microscope (SEM). Samples for this technique were ground with SiC papers up to 4000 grit, followed by \qty{20}{\minute} polishing with colloidal silica solution OPS. Final surface preparation for SEM was performed by ion beam milling using the Leica EM RES102 system. The images were taken using Trinity in-lens detectors in opti-plan mode with acceleration voltage of \qty{10}{\kilo\volt} and \qty{0.8}{\nano\ampere} beam current. Crystallographic orientation and microstructure were analysed by electron backscatter diffraction (EBSD) using the same specimens as for the conventional SEM imaging, using a \qty{10}{\kilo\volt} accelerating voltage and beam currents between \qty{13}{\nano\ampere} and \qty{51}{\nano\ampere}. The current used was selected according to the grain size at each deformation stage. EBSD datasets were processed and cleaned using the MTEX toolbox in MATLAB \cite{Bachmann2010-fa}.

Phase analysis was performed by X-ray diffraction (XRD) in Bragg--Brentano geometry using a Rigaku MiniFlex diffractometer with Cu K$\alpha$ radiation. Cut and polished samples, a few hundred micrometres thick, were used. The sample was azimuthally rotated at \qty{80}{\rpm} during exposure to improve grain statistics. The resulting diffraction profiles were analysed by Rietveld refinement using GSAS-II software.

Crystallographic texture was characterised by XRD using the Schulz reflection method, described in \cite{Schulz1949-jb,Chateigner1992-qc}, on a PHILIPS X’Pert PRO MRD diffractometer. All measurements were performed on samples cut perpendicular to the swaging direction (see Figure~\ref{fig: swaging setup diagram}b). For the hcp phase, the  reflections from planes $\{10\bar{1}0\}$, $\{0002\}$, $\{10\bar{1}1\}$, $\{10\bar{1}2\}$, $\{11\bar{2}0\}$ and $\{11\bar{2}2\}$ were collected, while for the bcc phase the $\{110\}$, $\{200\}$ and $\{211\}$ reflections were measured. The resulting pole figures were processed using MTEX toolbox, applying background subtraction and defocus compensation based on experimentally recorded background data. Because of the defocusing effect, only pole-figure regions of angle $\chi$ between \qty{0}{\degree} and \qty{75}{\degree} were used for subsequent analysis. The corrected pole figures were used to reconstruct the orientation distribution function (ODF) using the Williams--Imhof--Matthies--Vinel algorithm \cite{Matthies1988-vv,Singh2019-qt}. The pole figures presented in this work were subsequently recalculated from the ODF to account for the limited angular coverage of the experimental measurements and to enable the representation of the bcc $\{111\}$ pole figure, which cannot be measured directly due to the systematic extinction of the corresponding reflections.

Phase transformations were examined by electric resistometry measurements using a four-point method in a custom-built apparatus at a heating rate of \qty{5}{\degreeCelsius\per\minute} under an argon atmosphere \cite{Zhanal2015-yd}. 

Compression testing was performed on an Instron 5582 using cylindrical samples of height \qty{7.5}{\milli\meter} and diameter \qty{5}{\milli\meter}.

Microhardness was measured on transverse sections of the swaged rods using a Qness Q10 hardness tester under a load of \qty{500}{\gram} (\unit{\HVhalf}). The indentation grid consisted of concentric circles separated by \qty{250}{\micro\meter}, with individual indentations positioned along each circle at an approximate spacing of \qty{250}{\micro\meter}.

Light microscopy was performed using metallographic light microscope Zeiss Axio Observer 7 on samples polished using SiC papers and OPS solution and then etched using HF:HNO$_{3}$:H$_{2}$O in 10:45:45 ratio.

The oxygen and nitrogen content of swaged alloys was investigated using carrier gas hot extraction (CGHE) with G8 Galileo instrument, calibrated with AR628 standard. 

The Python and MATLAB scripts used for processing crystallographic texture and compression test data were developed with the assistance of ChatGPT (OpenAI) for code development and debugging. All scripts and resulting calculations were reviewed and validated by the authors.


\section{Results}

\subsection{Initial state of the material and the $\alpha \rightarrow \beta$ transformation}
\begin{figure}
    \centering
    \includegraphics[width=1\linewidth]{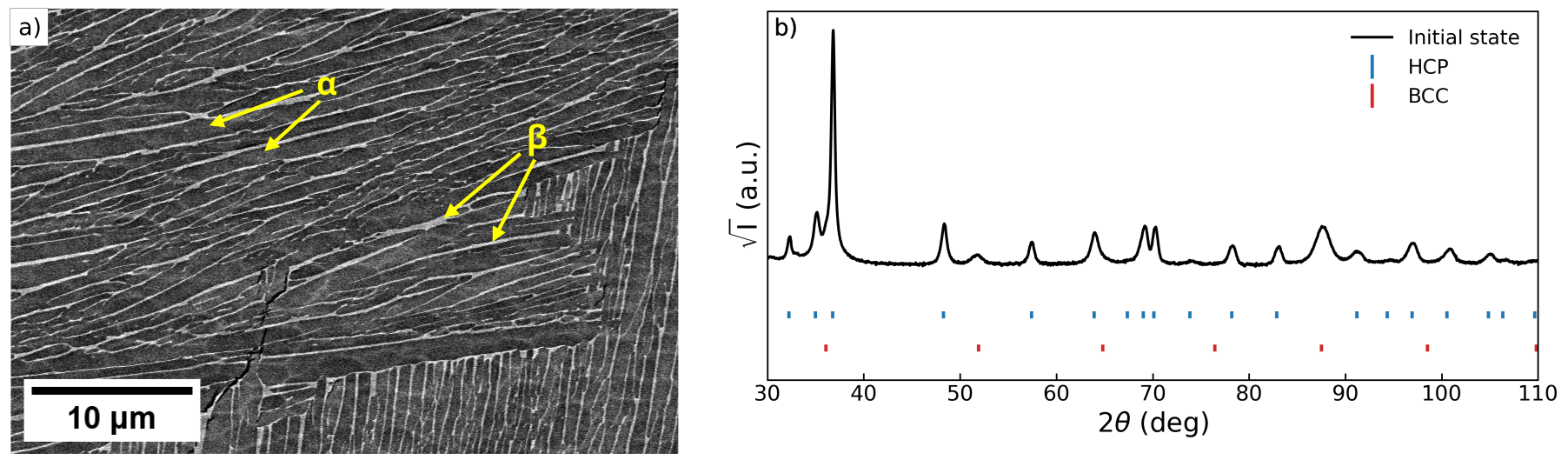}
    \caption{The initial cast + homogenised condition. a) SEM image, b) XRD line profile}
    \label{fig: Initial SEM + XRD}
\end{figure}

Before the hot swaging, the microstructure and phase composition of the cast, homogenised and furnace cooled condition were investigated as well as the phase transformations occurring in the material. In this state the material exhibited a two-phase $\alpha+\beta$ colony-type lamellar microstructure as shown in Figure~\ref{fig: Initial SEM + XRD}a. The phase fractions were estimated from the XRD measurement (see Figure~\ref{fig: Initial SEM + XRD}b) to be approximately \qty{15}{\wtpercent} bcc ($\beta$) phase and \qty{85}{\wtpercent} hcp ($\alpha$) phase. 

\begin{figure}
    \centering
    \includegraphics[width=0.5\linewidth] {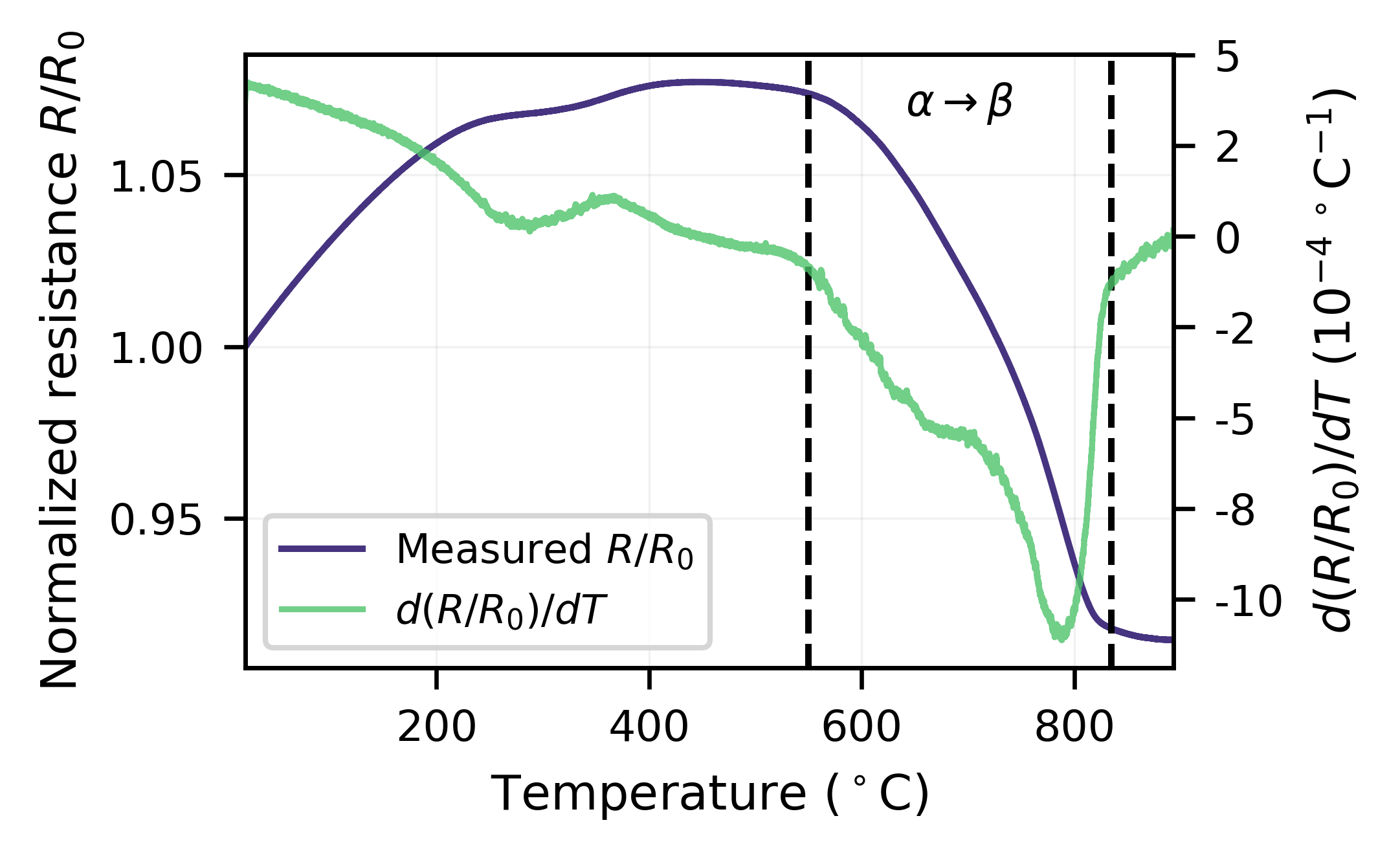}
    \caption{Electrical resistivity measurement showing the measured values of normalised electrical resistance and the first derivative of resistance with respect to temperature. The highlighted region between 550 and 830 $\degree$C represents the temperature range in which the $\alpha \rightarrow \beta$ transformation occurs.}
    \label{fig:Resisto}
\end{figure}

To determine the temperature ranges corresponding to the $\beta$ and $\alpha+\beta$ phase fields, phase transformations were investigated by electrical resistivity measurements. The normalised electrical resistance and its temperature derivative as functions of temperature are shown in Figure~\ref{fig:Resisto}. Prior to calculating the derivative, the normalised resistance data were smoothed using a Savitzky--Golay filter with a third-order polynomial and a smoothing window of \qty{3}{\degreeCelsius}, corresponding to 150 data points.

The $\alpha \rightarrow \beta$ transformation was found to begin at approximately \qty{550}{\degreeCelsius} and proceed to the $\beta$-transus temperature, which was determined to be approximately \qty{830}{\degreeCelsius}. From these results the temperature for the hot swaging in the $\alpha+\beta$ phase region was determined to be \qty{780}{\degreeCelsius}, \qty{50}{\degreeCelsius} below the measured $\beta$-transus. The temperature for swaging in the $\beta$ phase region was determined in the same fashion to be \qty{880}{\degreeCelsius}.
\newpage
\subsection{Microstructural evolution during hot rotary swaging}

Following the initial characterisation, the alloy was subjected to hot rotary swaging according to the processing route shown in Figure~\ref{Swaging process graph}. The first deformation stage was carried out in the $\beta$ phase region at \qty{880}{\degreeCelsius}, as explained in Section~\ref{section: rod swaging}. Subsequent swaging was performed in the $\alpha+\beta$ phase field at \qty{780}{\degreeCelsius}, resulting in a further diameter reduction to \qty{6.3}{\milli\meter}.

\begin{figure}[h!]
    \centering
    \includegraphics[
        width=\textwidth,
        height=0.85\textheight,
        keepaspectratio
    ]{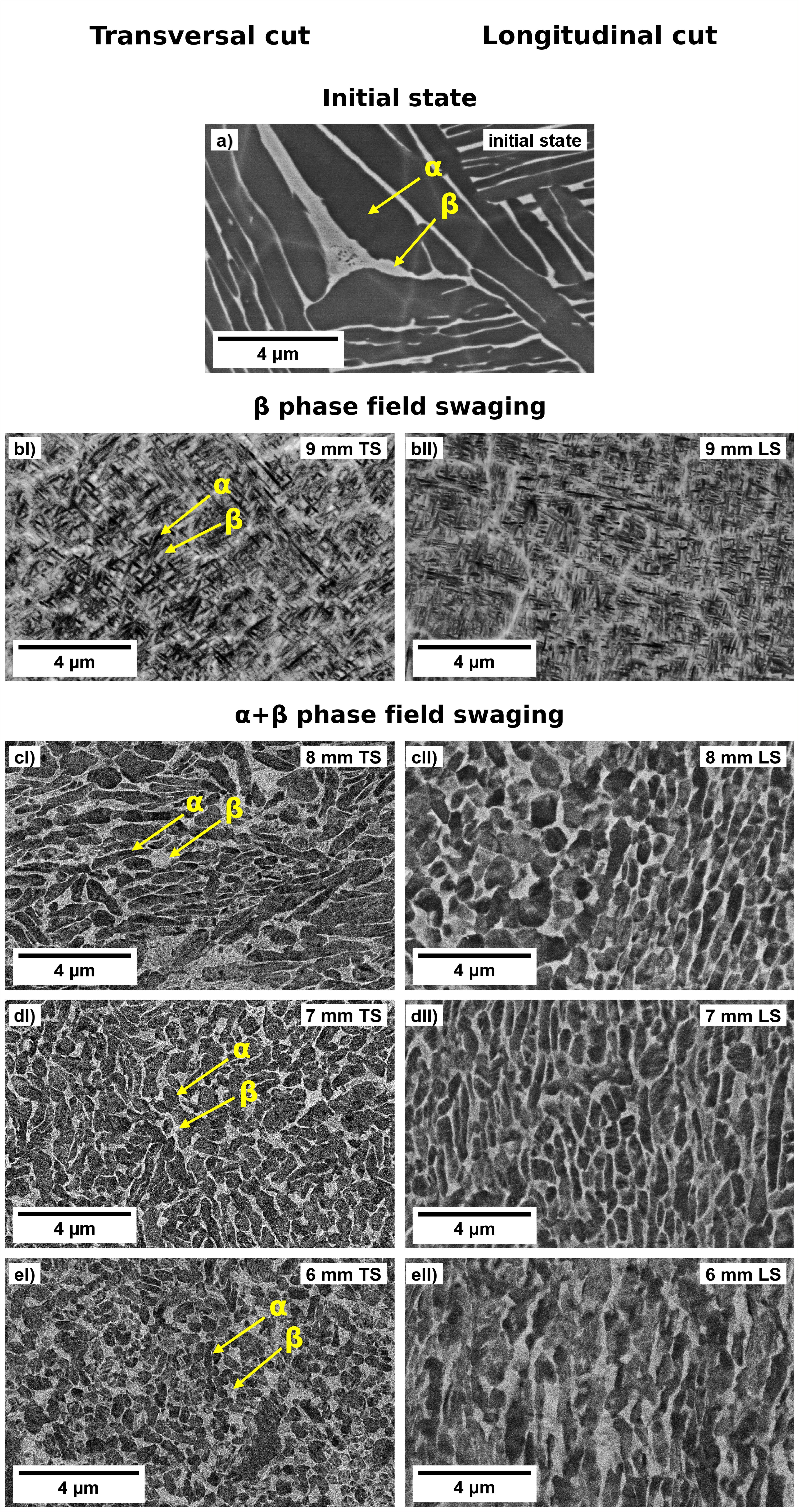}
    \caption{Evolution of the microstructure after sequential hot-swaging steps. a) The initial state after homogenisation contains relatively large lamellar $\alpha$ phase. b) Rapid cooling after swaging in the $\beta$ phase region produces a fine Widmanstätten $\alpha$ microstructure. c-e) After swaging in the $\alpha+\beta$ phase region, the material exhibits progressive globularisation of the $\alpha$ phase.}
    \label{SEM figure grid}
\end{figure}

\begin{figure}[!htbp]
    \centering

    \includegraphics[width=0.88\linewidth]{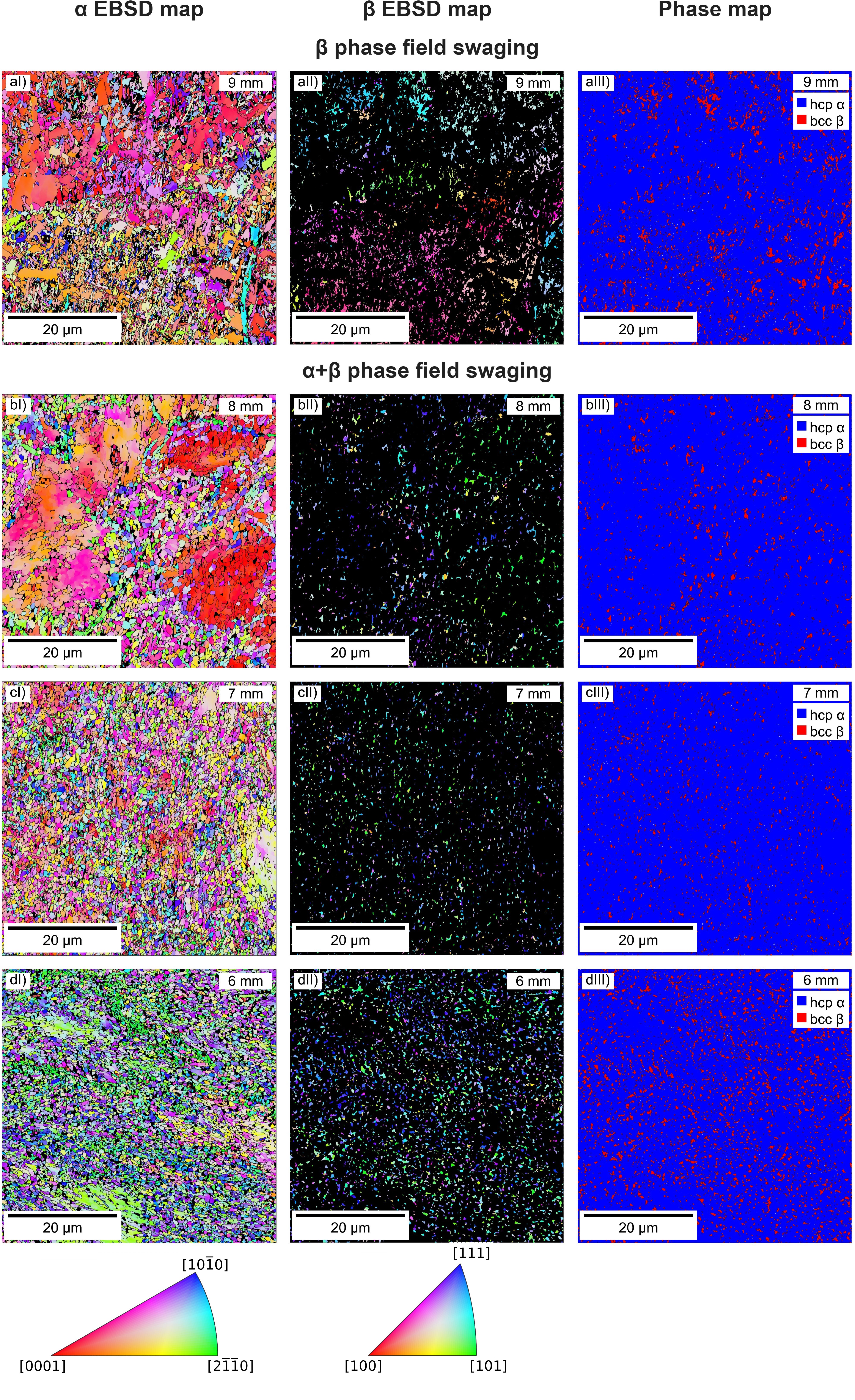}
    \caption{EBSD maps of individual steps of swaging performed on transverse samples. a) During swaging in the $\beta$ phase field $\alpha$ colony microstructure is formed and large $\beta$ grains are retained. b-d) After swaging in the $\alpha+ \beta$ phase field.}
    \label{fig:ebsd-grid}
\end{figure}

The microstructural evolution induced by the swaging process was examined by SEM on both the transverse and longitudinal samples, see Figure~\ref{SEM figure grid}. In addition, EBSD measurements were conducted to evaluate changes in grain morphology and crystallographic texture resulting from the applied thermomechanical deformation, shown in Figure~\ref{fig:ebsd-grid}. The EBSD measurement was performed only on the transverse samples.

\begin{figure}[h]
    \centering
    \includegraphics[width=0.28\linewidth]{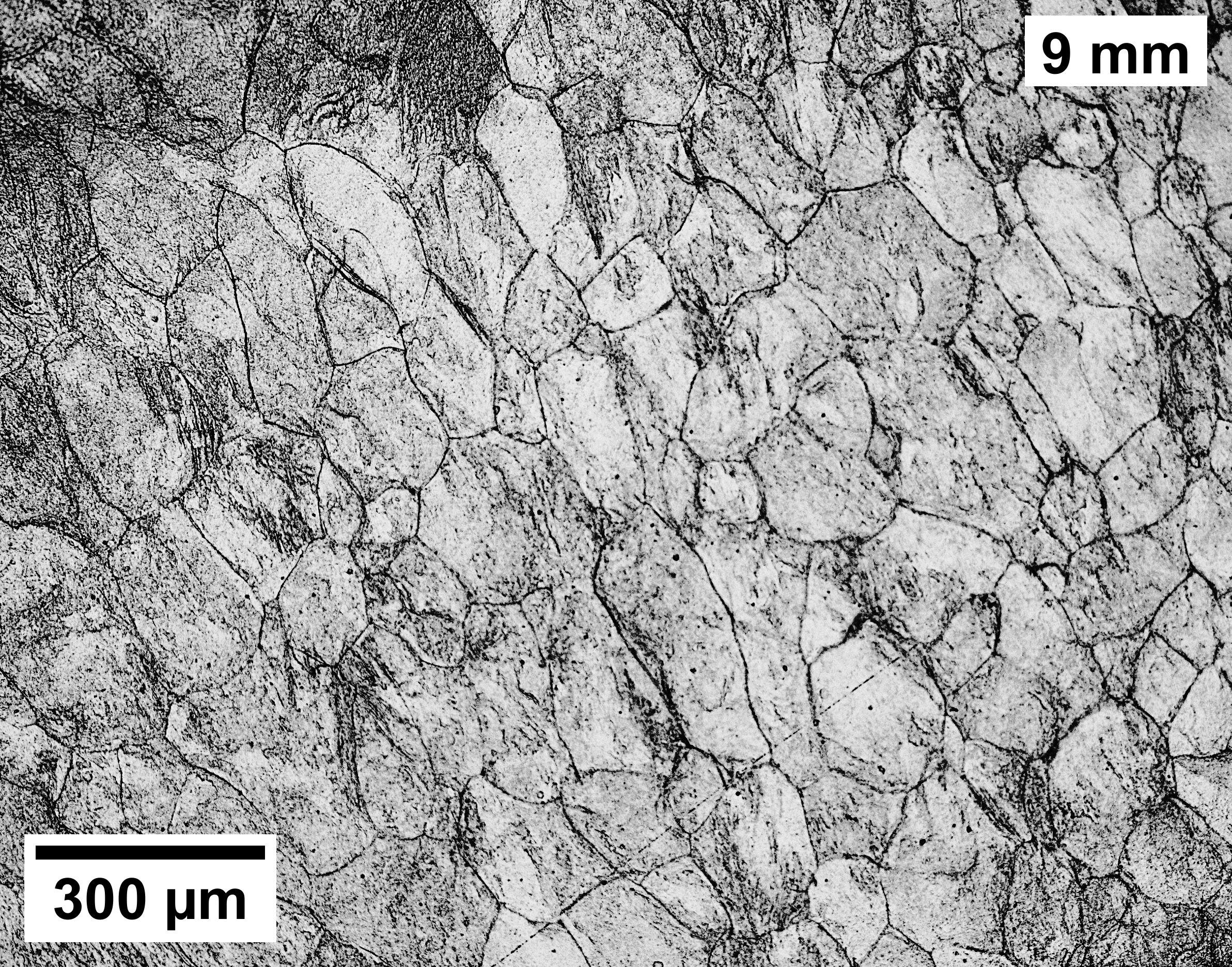}
    \caption{Light microscope image of etched $\beta$ grains after swaging in the $\beta$ phase field.}
    \label{fig: LM}
\end{figure}

Dynamic recrystallisation of the $\beta$ grains occurred during swaging above the $\beta$-transus temperature. During the processing, the prior-$\beta$ grain size was refined from the millimetre scale to approximately 100~$\mu$m, as seen from the LM image in Figure~\ref{fig: LM}. Subsequent rapid cooling from the $\beta$ phase region resulted in a microstructure consisting of retained $\beta$ phase and fine $\alpha$ lamellae, as shown in Figure~\ref{SEM figure grid}bI-II, where $\alpha$ lamellae are arranged into colonies with a common crystallographic orientation. Due to their low thickness, the individual $\alpha$ lamellae could not be resolved by EBSD. Consequently, the large indexed $\alpha$ grains in Figure~\ref{fig:ebsd-grid}aI represent $\alpha$ colonies rather than individual $\alpha$ lamellae.

The microstructure after deformation in the $\alpha+\beta$ phase field can be seen in the SEM images (Figures \ref{SEM figure grid}c-e) and EBSD maps (Figure~\ref{fig:ebsd-grid}b-d). During deformation in this phase field, the $\beta$ grains were refined further, as shown in Figure~\ref{fig:ebsd-grid}dII.

After the first swaging step in the $\alpha+\beta$ phase field (Figure~\ref{SEM figure grid}cI-II), new globularised $\alpha$ grains were observed, while remnants of the original $\alpha$ colonies were still retained. As discussed above, these colonies again appear as large $\alpha$ grains in Figure~\ref{fig:ebsd-grid}bI, as the constituent $\alpha$ lamellae could not be resolved individually by EBSD. 

In the final swaged condition (Figure~\ref{SEM figure grid}eI-II), the microstructure consisted predominantly of globularised $\alpha$ grains. However, small residual colonies originating from the initial lamellar $\alpha$ microstructure remained detectable even after the last swaging step. The final $\alpha$ grains are elongated along the swaging direction, a characteristic microstructural feature of rotary swaging that has been reported in commercially pure Ti and $\alpha+\beta$ titanium alloys~\cite{Palan2017,Dyakonov2023-ql}. SEM images \ref{SEM figure grid}cII,dII and eII suggest that the $\alpha$ grain elongation may become progressively more pronounced with successive swaging steps.

\newpage
\subsection{Texture changes during the swaging process}

The crystallographic texture evolution during swaging was analysed using both electron backscatter diffraction (EBSD) and X-ray diffraction (XRD). EBSD provided local information on crystallographic orientations in the centre of the swaged samples, as shown in Figure~\ref{fig:ebsd-grid}. The corresponding pole figures (PFs) calculated from these maps are presented in supplementary material in supplementary Figure~\ref{EBSD PF grid}. XRD measurements performed using the Schulz reflection method provided texture information averaged over the analysed sample area. The XRD results are presented as inverse pole figures (IPFs) along the swaging direction in Figure~\ref{XRD IPFs}, while the recalculated PFs are shown in supplementary Figure~\ref{XRD texture}. In both cases, the measurements were performed on transverse sections with the analysed surface perpendicular to the swaging direction (SD), as illustrated in Figure~\ref{fig: swaging setup diagram}b.

After the high-temperature swaging above the $\beta$-transus, EBSD did not reveal a well-defined crystallographic texture. This was mainly due to the large $\beta$ grain size relative to the EBSD scan area of \qty{50}{\micro\meter} $\times$ \qty{50}{\micro\meter}, so only a limited number of strong orientations were captured in the $\beta$ pole figures (supplementary Figure~\ref{EBSD PF grid}). These orientations were inherited by the $\alpha$ phase, as indicated by the similarities between the $(0002)_{\mathrm{hcp}}$ and $(110)_{\mathrm{bcc}}$ pole figures, and between the $(11\bar20)_{\mathrm{hcp}}$ and $(111)_{\mathrm{bcc}}$ pole figures, shown in supplementary Figure~\ref{EBSD PF grid}. This inheritance is consistent with the Burgers orientation relationship, $(0002)_{\mathrm{hcp}} \parallel (110)_{\mathrm{bcc}}$ and $\langle 11\bar20\rangle_{hcp} \parallel \langle 111\rangle_{bcc}$, which governs the $\beta \rightarrow \alpha$ transformation.

\begin{figure}[hb!]
    \centering
    \includegraphics[width=0.5\linewidth]{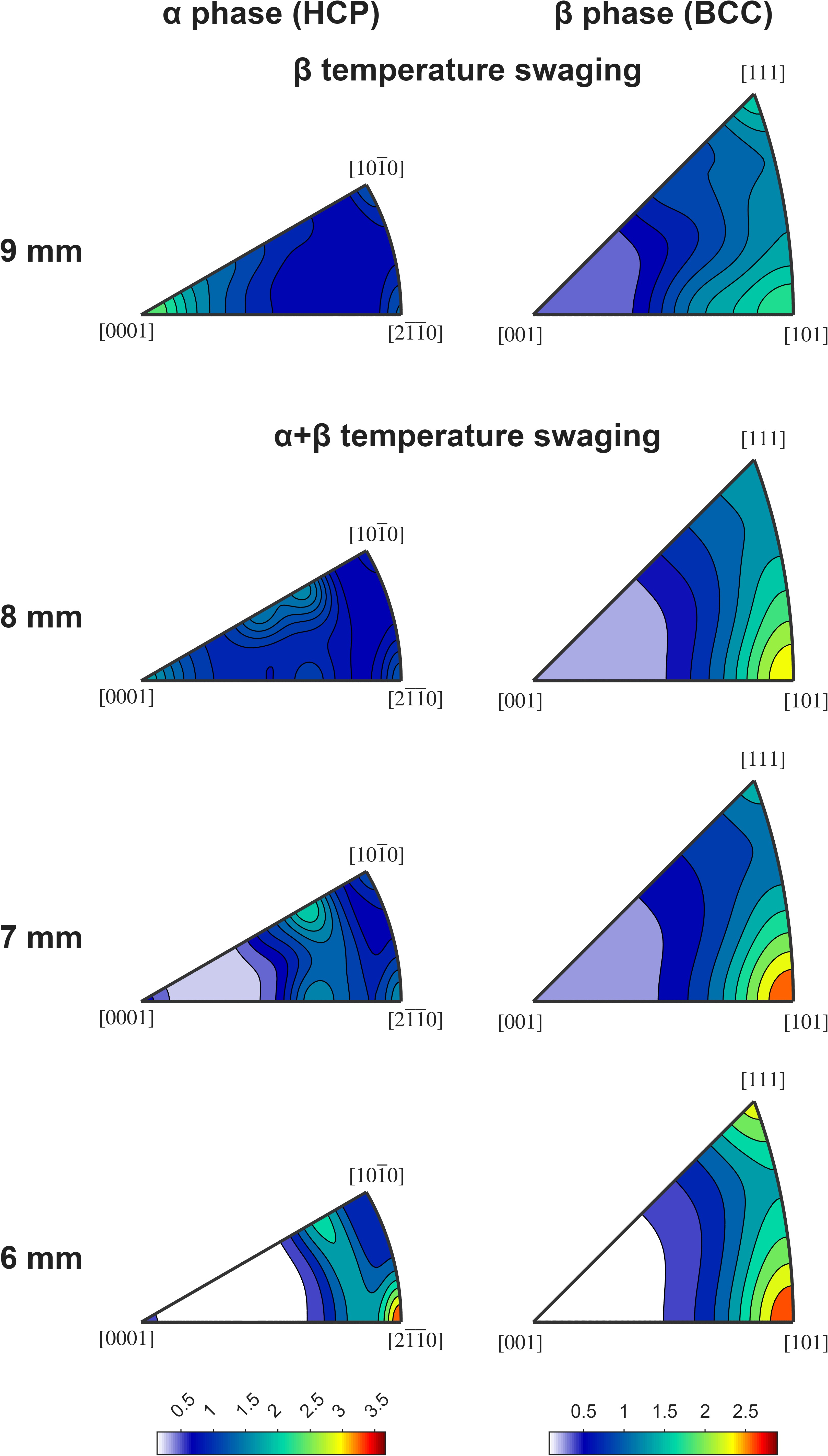}
    \caption{Inverse pole figure results from XRD texture measurement using Schulz method}
    \label{XRD IPFs}
\end{figure}

In contrast to the local EBSD observations, the XRD texture data show that a preferred texture developed in the swaged rod as a whole after high-temperature swaging. The inverse pole figures in Figure~\ref{XRD IPFs} indicate the formation of $\langle 110 \rangle_{\mathrm{bcc}}$ and $\langle 111 \rangle_{\mathrm{bcc}}$ texture components parallel to SD. The inherited $\langle 0001 \rangle_{\mathrm{hcp}} \parallel SD$ texture was also observed. A detailed view of the recalculated PFs is provided in supplementary Figure~\ref{XRD texture}.

After swaging in the intermediate $\alpha+\beta$ temperature range, the EBSD measurements (Figure~\ref{fig:ebsd-grid}) revealed a gradual tilting of the $\alpha$ basal planes, $(0002)_{\mathrm{hcp}}$, away from the swaging direction with increasing deformation. In the $\beta$ phase, the normals of the $(111)_{\mathrm{bcc}}$ planes became aligned with the swaging direction. This texture component was already present after the first swaging step in the $\alpha+\beta$ phase field and persisted up to the final deformation step.

The XRD texture measurements after $\alpha+\beta$ field swaging (Figure~\ref{XRD IPFs}) show the gradual development of a $\langle 11\bar{2}0\rangle_{\mathrm{hcp}} \parallel \mathrm{SD}$ fibre texture in the $\alpha$ phase. This is geometrically consistent with the progressive tilting of the $(0002){_\mathrm{hcp}}$ basal planes observed by EBSD. The $\langle 11\bar{2}0\rangle{_\mathrm{hcp}} \parallel \mathrm{SD}$ fibre texture became clearly established after the final swaging step to a diameter of \qty{6.3}{\milli\meter}. All $\alpha+\beta$ swaged states also exhibited a $\langle 10\bar{1}1\rangle_{\mathrm{hcp}} \parallel \mathrm{SD}$ texture component. In the $\beta$ phase, a $\langle 110\rangle_{\mathrm{bcc}} \parallel \mathrm{SD}$ texture was present throughout the intermediate-temperature swaging sequence and strengthened with subsequent swaging. After the final swaging step, a noticeable $\langle 111\rangle_{\mathrm{bcc}} \parallel \mathrm{SD}$ texture component was also observed.

\subsection{Phase composition evolution}
\begin{figure}
    \centering
    \includegraphics[width=0.9\linewidth]{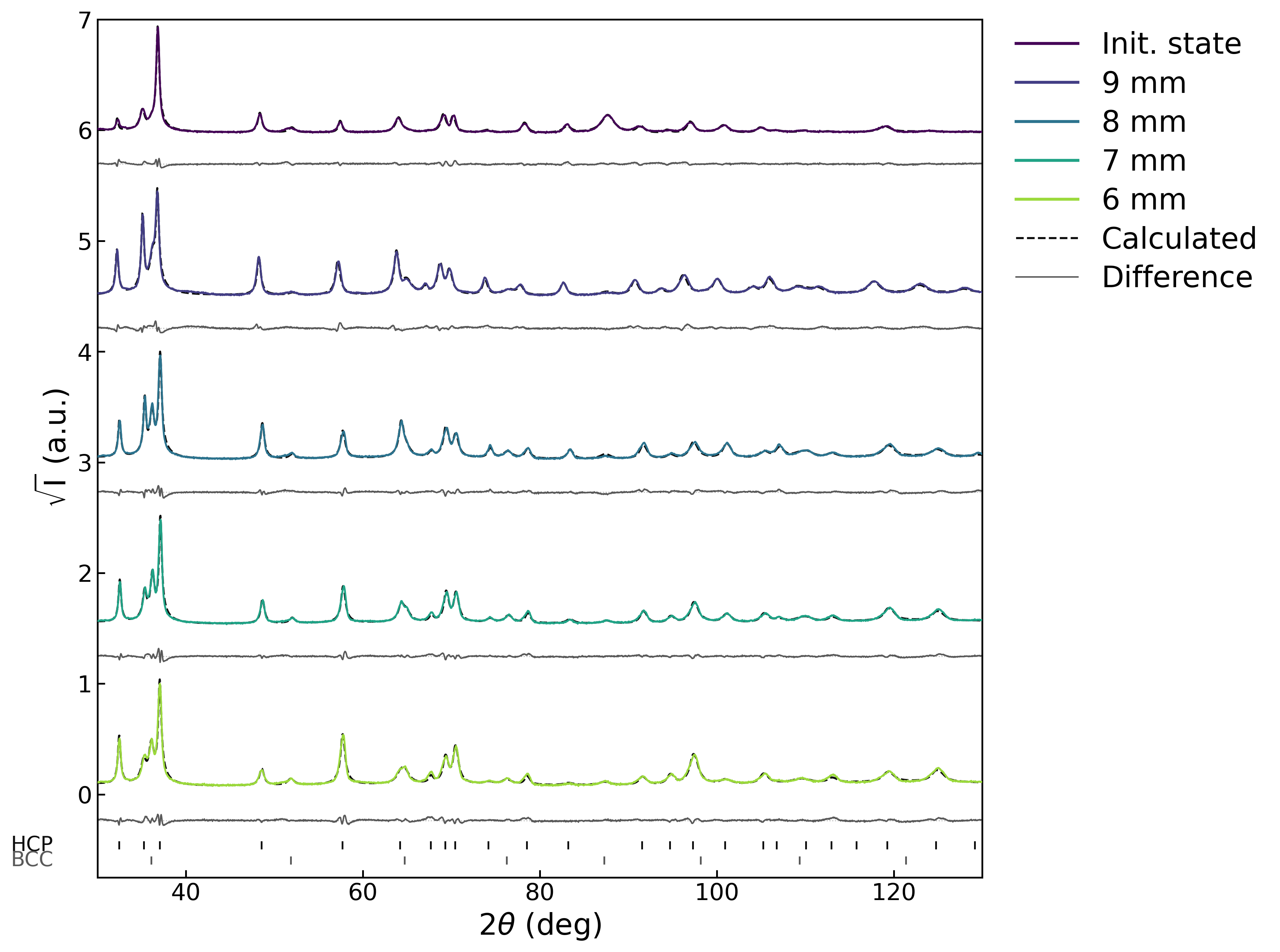}
\caption{Measured XRD profiles of the investigated states, together with the corresponding calculated profiles and difference curves obtained using the no-partition (NP) model. The positions of the $\alpha$-hcp and $\beta$-bcc reflections determined by Rietveld refinement are indicated below the diffraction profiles.}

    \label{fig: XRD BB}
\end{figure}

The evolution of the phase composition during rotary swaging was investigated by X-ray diffraction (XRD) in Bragg--Brentano geometry. The corresponding diffraction patterns are shown in Figure~\ref{fig: XRD BB}. Rietveld refinement was used to determine the phase weight fractions, summarized in Table~\ref{XRD phase composition results}, and the corresponding lattice parameters, listed in Table~\ref{tab:lattice_parameters}. The refinement quality metrics are provided in supplementary Table~\ref{tab:rietveld_fit_quality}.

For the quantitative phase analysis, two structural models were considered. In the first model, elemental partitioning between the phases was neglected, and both the hcp and bcc phases were therefore assumed to have the nominal alloy composition (labelled NP - no partition). In the second, self-consistent model, complete partitioning of the alloying elements was assumed, with Al assigned exclusively to the hcp phase and Nb exclusively to the bcc phase, resulting in Zr--Al and Zr--Nb compositions for the hcp and bcc phases, respectively  (labelled FP - full partition). The final proposed phase composition was then taken as a mean of these two extreme cases. Figure~\ref{fig: XRD BB} presents the measured diffraction profiles together with the corresponding fits obtained using the NP model and the associated difference curves.

The whole-pattern fit quality was comparable across all investigated states, with $R_{\mathrm{wp}}$ values of \qtyrange{11.44}{12.78}{\percent} and GOF values of 4.36--5.73. These values indicate that the principal diffraction features and overall intensity distributions were reproduced reasonably well, although the GOF values substantially above unity show that the residual deviations exceed those expected from counting statistics alone. In addition, the Durbin--Watson statistics were very low, ranging from 0.167 to 0.288, indicating strong positive correlation between neighbouring residuals and therefore the presence of systematic rather than purely random discrepancies. These residuals are attributed mainly to pronounced crystallographic texture, partial reflection overlap, and complex asymmetric peak shapes that cannot be fully reproduced by the employed profile functions. Such peak asymmetry has previously been observed in deformed Zr alloys; Sarkar et al.~\cite{Sarkar2008-ak} reported asymmetric diffraction peaks in deformed Zircaloy-2 and Zr--2.5Nb alloys and related them to deformation-induced microstructural inhomogeneity.

Consequently, although the refinements are considered adequate for approximate quantitative phase analysis, the formal Rietveld uncertainties of approximately \qtyrange{0.2}{0.5}{\wtpercent} reflect only the statistical precision of the fitted models and do not account for other sources of systematic and model-dependent uncertainty. The true analytical uncertainty is therefore expected to be higher. Considering the variation between the two structural models together with these additional sources of uncertainty, a more conservative uncertainty of \qty{1}{\wtpercent} was assigned to the proposed phase composition.

 Only the hcp $\alpha$ and bcc $\beta$ phases were detected in all stages of the swaging process. No intermetallic phases were identified, although Thermo-Calc predicts the intermetallic phase AlZr$_{2}$ to be present in this alloy under equilibrium conditions (Figure~\ref{Thermo calc figure}).
A substantial fraction of the $\beta$ phase, approximately \qtyrange{12}{15}{\wtpercent}, was retained throughout the thermomechanical processing, while the hcp phase accounted for approximately \qtyrange{85}{88}{\wtpercent} of the microstructure (Table~\ref{XRD phase composition results}). Despite the conservative uncertainty of \qty{1}{\wtpercent} assigned to the proposed phase fractions, a gradual increase in the hcp fraction is apparent with successive swaging steps, from $85\pm1$\,\unit{\wtpercent} in the initial state to $88\pm1$\,\unit{\wtpercent} in the final \qty{6}{\milli\meter} state. Correspondingly, the retained bcc fraction decreases from approximately $15\pm1$ to $12\pm1$\,\unit{\wtpercent}.

\begin{table}[htbp]
\centering
\resizebox{\textwidth}{!}{%
\begin{tabular}{|c|c|c|c|c|c|c|} \hline
State
& hcp-NP (wt.\%)
& bcc-NP (wt.\%)
& hcp-FP (wt.\%)
& bcc-FP (wt.\%)
& Proposed hcp (wt.\%)
& Proposed bcc (wt.\%) \\ \hline

Initial
& $84.8 \pm 0.2$
& $15.2 \pm 0.2$
& $85.3 \pm 0.2$
& $14.7 \pm 0.2$
& $85.0 \pm 1.0$
& $15.0 \pm 1.0$ \\

9 mm
& $84.9 \pm 0.3$
& $15.1 \pm 0.3$
& $85.7 \pm 0.3$
& $14.3 \pm 0.3$
& $85.3 \pm 1.0$
& $14.7 \pm 1.0$ \\

8 mm
& $85.8 \pm 0.3$
& $14.3 \pm 0.3$
& $86.5 \pm 0.3$
& $13.5 \pm 0.3$
& $86.1 \pm 1.0$
& $13.9 \pm 1.0$ \\

7 mm
& $87.0 \pm 0.4$
& $13.0 \pm 0.4$
& $87.7 \pm 0.4$
& $12.3 \pm 0.4$
& $87.3 \pm 1.0$
& $12.7 \pm 1.0$ \\

6 mm
& $87.7 \pm 0.5$
& $12.3 \pm 0.5$
& $88.3 \pm 0.5$
& $11.7 \pm 0.5$
& $88.0 \pm 1.0$
& $12.0 \pm 1.0$ \\ \hline

\end{tabular}%
}
\caption{Phase fractions obtained from Rietveld refinement using the no-partition (NP) and full-partition (FP) models, including the corresponding fit uncertainties. The proposed phase fractions were calculated from the two models and are reported with an estimated uncertainty of $\pm 1$~wt.\%.}
\label{XRD phase composition results}
\end{table}
\begin{table}[h!]
\centering

\begin{tabular}{|c|c|c|c|c|c|c|}
\hline
& \multicolumn{3}{c|}{No partition (NP)} & \multicolumn{3}{c|}{Full partition (FP)} \\ \hline
State & $a_{\alpha}$ (\AA) & $c_{\alpha}$ (\AA) & $a_{\beta}$ (\AA)
& $a_{\alpha}$ (\AA) & $c_{\alpha}$ (\AA) & $a_{\beta}$ (\AA) \\ \hline
Initial & 3.2108(4) & 5.1228(5) & 3.5219(3) & 3.2108(4) & 5.1228(5) & 3.5219(3) \\
9 mm & 3.2361(3) & 5.1507(2) & 3.5305(8) & 3.2361(3) & 5.1507(2) & 3.5305(8) \\
8 mm & 3.2058(3) & 5.1131(3) & 3.5354(6) & 3.2058(3) & 5.1131(3) & 3.5353(6) \\
7 mm & 3.2042(3) & 5.1156(4) & 3.5325(5) & 3.2042(3) & 5.1156(4) & 3.5324(5) \\
6 mm & 3.2039(3) & 5.1216(7) & 3.5382(7) & 3.2039(3) & 5.1216(7) & 3.5382(7) \\
\hline
\end{tabular}

\vspace{0.8em}

\begin{tabular}{|c|c|c|c|}
\hline
State
& Proposed $a_{\mathrm{\alpha}}$ (\AA)
& Proposed $c_{\mathrm{\alpha}}$ (\AA)
& Proposed $a_{\mathrm{\beta}}$ (\AA) \\
\hline
Initial
& $3.211 \pm 0.001$
& $5.123 \pm 0.001$
& $3.522 \pm 0.001$ \\
9 mm
& $3.236 \pm 0.001$
& $5.151 \pm 0.001$
& $3.530 \pm 0.001$ \\
8 mm
& $3.206 \pm 0.001$
& $5.113 \pm 0.001$
& $3.535 \pm 0.001$ \\
7 mm
& $3.204 \pm 0.001$
& $5.116 \pm 0.001$
& $3.532 \pm 0.001$ \\
6 mm
& $3.204 \pm 0.001$
& $5.122 \pm 0.001$
& $3.538 \pm 0.001$ \\
\hline

\end{tabular}

\caption{Lattice parameters obtained by Rietveld refinement using the no-partition (NP) and full-partition (FP) models, including the corresponding fit uncertainties. The proposed lattice parameters were calculated as the mean of the values obtained from the two models.}
\label{tab:lattice_parameters}
\end{table}
The fitted lattice parameters are summarized in Table~\ref{tab:lattice_parameters}. The NP and FP models yielded nearly identical values, generally differing only in the fifth decimal place. The proposed lattice parameters were calculated as the arithmetic means of the values obtained from the two models. Since potential systematic errors were not explicitly corrected or quantified, a conservative estimated uncertainty of \qty{0.001}{\angstrom} was assigned to the proposed values.

The initial state and the samples swaged in the $\alpha+\beta$ phase region (8, 7, and \qty{6}{\milli\meter}) exhibit similar $\alpha$-phase lattice parameters, with $a_{\alpha}$ ranging from \qty{3.204}{\angstrom} to \qty{3.211}{\angstrom} and $c_{\alpha}$ from \qty{5.113}{\angstrom} to \qty{5.123}{\angstrom}. These values are lower than those of unalloyed $\alpha$-Zr, for which $a_{\alpha}=\qty{3.233}{\angstrom}$ and $c_{\alpha}=\qty{5.147}{\angstrom}$~\cite{Liu2021ZrPhaseTransition}. The reduced lattice parameters are consistent with the contraction of the $\alpha$-Zr lattice caused by Al addition, as reported for binary Zr--Al alloys by Tiwari and Tangri~\cite{Tiwari1970}. In contrast, the state swaged above the $\beta$-transus (\qty{9}{\milli\meter}) exhibits distinctly larger lattice parameters of $a_{\alpha}=\qty{3.236(1)}{\angstrom}$ and $c_{\alpha}=\qty{5.151(1)}{\angstrom}$. This expansion after the initial high-temperature swaging step cannot be explained simply by Al depletion, since rapid quenching from a chemically homogeneous $\beta$ phase should retain approximately the bulk Al concentration of \qty{6.46}{\atpercent}. A higher Nb concentration is also unlikely to explain the expansion, as increasing Nb content has been reported to decrease the lattice parameters of $\alpha$-Zr.

A potentially more significant contribution to the observed lattice expansion may arise from residual stresses associated with the $\beta \rightarrow \alpha$ transformation. Rapid cooling of Zr--Nb alloys from the $\beta$ field can produce hcp $\alpha'$ martensite \cite{CCT_ZrNbO}. However, reported $\alpha'$ lattice parameters ($a_{\alpha'}=\qty{3.211}{\angstrom}$, $c_{\alpha'}=\qty{5.115}{\angstrom}$) are considerably smaller than those measured in the present \qty{9}{\milli\meter} state \cite{Srivastava2004}. Thus, martensite formation alone cannot explain the observed lattice expansion, although transformation-induced residual and intergranular stresses may contribute. The possible martensitic transformation is further supported by the fine needle-like $\alpha$ morphology observed after high-temperature swaging (Figure~\ref{SEM figure grid}bI--II), consistent with reported martensitic morphologies in dilute Zr--Nb alloys \cite{Srivastava2000}.

An effect of interstitial oxygen or nitrogen on the lattice parameters can be excluded in the present case, as CGHE showed no measurable bulk uptake of either element during the swaging process (Table~\ref{CGHE data}). The larger lattice parameters of the \qty{9}{\milli\meter} state are therefore most plausibly associated with the non-equilibrium $\beta \rightarrow \alpha$ transformation during cooling, potentially involving both changes in Al partitioning and transformation-induced residual strain. Further phase-specific chemical analysis or stress-resolved diffraction measurements would be required to separate the relative contributions of these two effects.

\begin{table}[h]
\begin{tabular}{|c|c|c|} \hline
State         & Oxygen content (\unit{\wtpercent}) & Nitrogen content (\unit{\wtpercent}) \\ \hline
Initial state & 0.025 ± 0.002          & 0.007 ± 0.002            \\
\qty{6}{\milli\meter} & 0.024 ± 0.002 & 0.006 ± 0.002 \\ \hline
\end{tabular}
\caption{CGHE results showing the oxygen and nitrogen contents in the initial state before swaging and in the final \qty{6}{\milli\meter} swaged state.}
\label{CGHE data}
\end{table}

\subsection{Compression tests}
\begin{figure}[t]
    \centering

    \begin{minipage}[c]{0.58\textwidth}
        \centering
        \includegraphics[width=\linewidth]{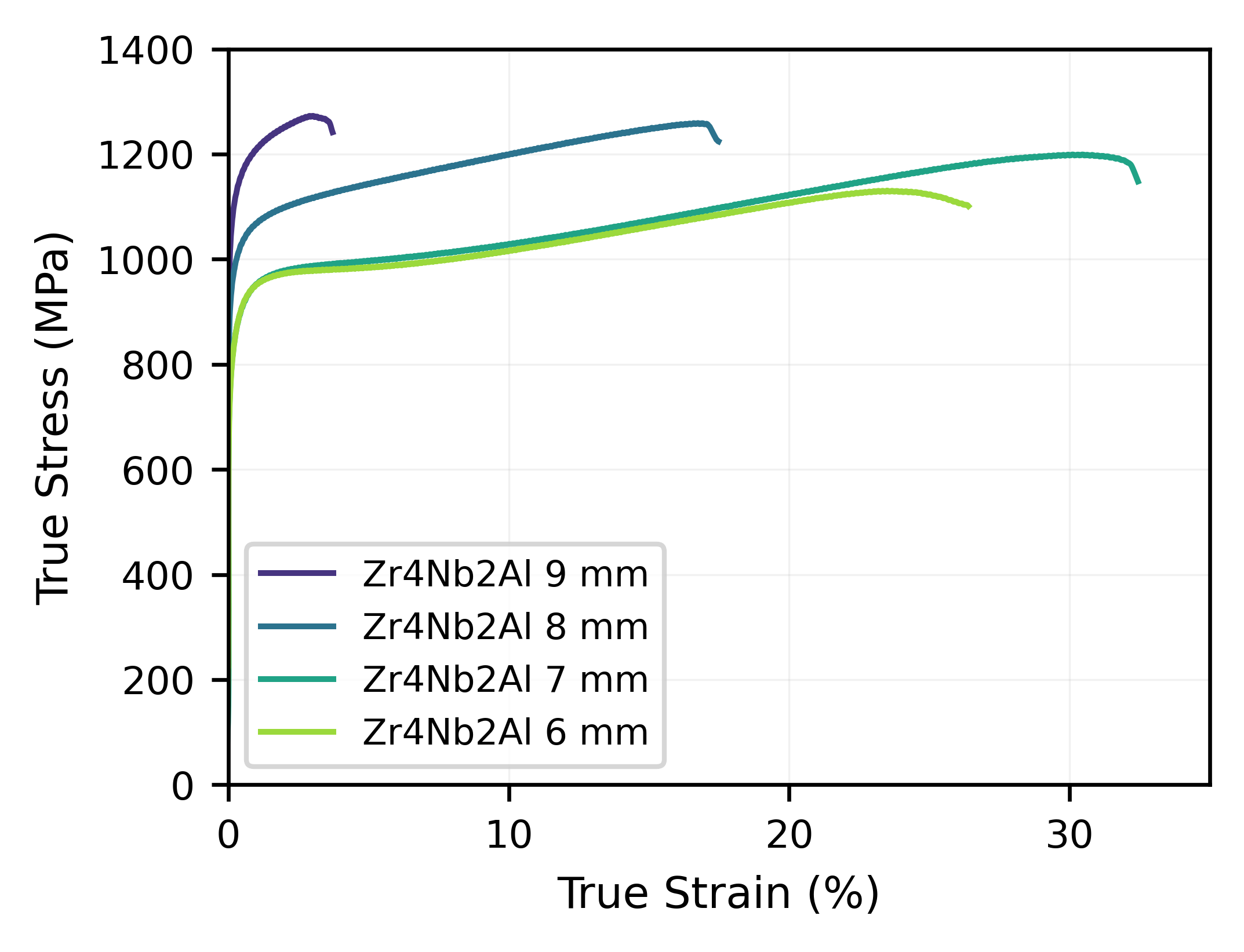}
        \captionof{figure}{Compression stress--strain curves of the individual swaged states.}
        \label{compression graph}
    \end{minipage}
    \hfill
\begin{minipage}[c]{0.4\textwidth}
    \centering
    \resizebox{\linewidth}{!}{%
        \begin{tabular}{|c|c|c|c|}
            \hline
            Sample & $\sigma_{0.2}$ (\unit{\mega\pascal}) & UCS (\unit{\mega\pascal}) & $\varepsilon_{f}$ (\unit{\percent}) \\
            \hline
            \qty{9}{\milli\meter} & 1103 & 1273 & 3.6  \\
            \qty{8}{\milli\meter} & 979  & 1259 & 17.1 \\
            \qty{7}{\milli\meter} & 839  & 1199 & 32.2 \\
            \qty{6}{\milli\meter} & 839  & 1131 & 26.4 \\
            \hline
        \end{tabular}%
    }
    \captionof{table}{Mechanical properties determined from the compression tests.}
    \label{compresion results}
\end{minipage}
\end{figure}

The results of ex-situ compression tests performed on the individual swaged states are shown in Figure~\ref{compression graph}. Only the plastic part of the stress--strain curves is presented. The determined \qty{0.2}{\percent} yield strength, $\sigma_{0.2}$, ultimate compressive strength (UCS), and strain at fracture ($\varepsilon_{f}$) are summarized in Table~\ref{compresion results}.

The mechanical response changes considerably with increasing swaging deformation. The most pronounced trend is the progressive decrease in yield strength, while the differences in UCS remain comparatively small. The \qty{9}{\milli\meter} state, produced by swaging in the $\beta$ field, exhibits the highest yield strength, with $\sigma_{0.2}=\qty{1103}{\mega\pascal}$, and reaches a UCS of \qty{1273}{\mega\pascal}. However, only limited post-yield deformation is observed, with fracture occurring at a strain of \qty{3.6}{\percent}.

Further swaging in the $\alpha+\beta$ field decreases the yield strength while increasing the extent of strain hardening and plastic deformation. For the \qty{8}{\milli\meter} state, $\sigma_{0.2}$ decreases to \qty{979}{\mega\pascal}, whereas the UCS remains close to that of the \qty{9}{\milli\meter} state at \qty{1259}{\mega\pascal}. Consequently, the stress increase between yielding and the maximum compressive stress becomes more pronounced. At the same time, the strain at fracture increases substantially to \qty{17.1}{\percent}.

A further decrease in yield strength is observed for the \qty{7}{\milli\meter} and \qty{6}{\milli\meter} states, both reaching $\sigma_{0.2}=\qty{839}{\mega\pascal}$. Despite this approximate \qty{24}{\percent} reduction in yield strength relative to the \qty{9}{\milli\meter} state, the UCS remains comparatively high, reaching \qty{1199}{\mega\pascal} for the \qty{7}{\milli\meter} state and \qty{1131}{\mega\pascal} for the \qty{6}{\milli\meter} state. The \qty{7}{\milli\meter} state therefore exhibits the largest post-yield increase in stress together with the highest strain at fracture of \qty{32.2}{\percent}. After the final swaging step, the yield strength remains unchanged, whereas the UCS decreases moderately and the strain at fracture is reduced to \qty{26.4}{\percent}. Overall, progressive swaging thus produces a substantial reduction in yield strength accompanied by an increased capacity for post-yield strain hardening and deformation, while the corresponding decrease in UCS is considerably less pronounced.

After compression testing, specimen barreling was assessed from the diameters measured at the specimen ends, \(D_\mathrm{end}\), and at mid-height, \(D_\mathrm{center}\). The relative barreling was characterised as \(B_D=(D_\mathrm{center}-D_\mathrm{end})/D_\mathrm{end}\). The resulting values were \qty{0.6}{\percent}, \qty{1.8}{\percent}, \qty{5.2}{\percent}, and \qty{4.4}{\percent} for the \qty{9}{\milli\meter}, \qty{8}{\milli\meter}, \qty{7}{\milli\meter}, and \qty{6}{\milli\meter} states, respectively. The increase in barreling for the latter states is associated with their substantially larger plastic strains prior to fracture. No friction correction was applied to the reported stress--strain curves; consequently, the stresses at large plastic strains should be regarded as uncorrected true stresses calculated assuming homogeneous, volume-conserving deformation.



\subsection{Microhardness evolution}
\begin{figure}[h!]
    \centering

    \includegraphics[width=0.55\linewidth]{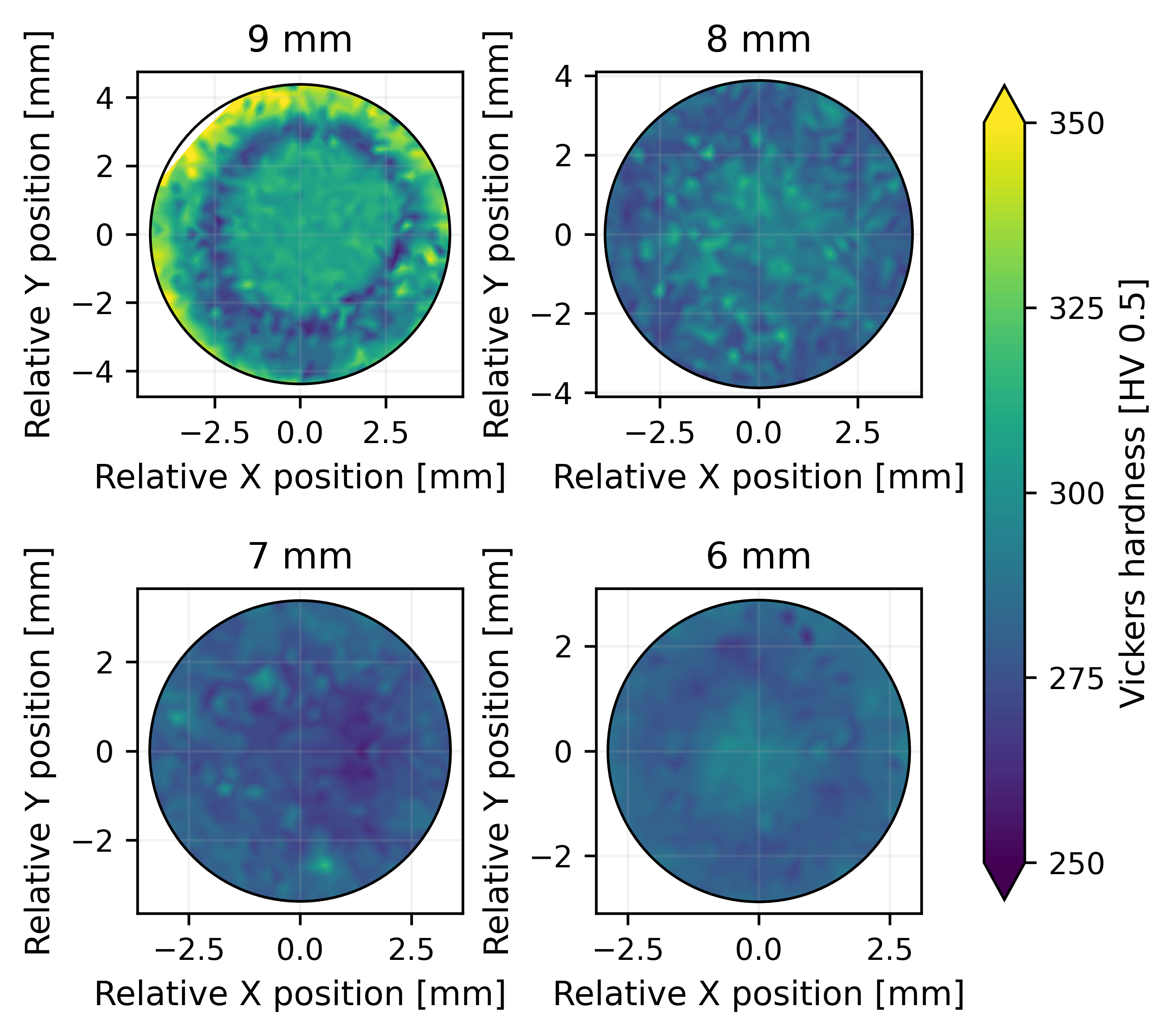}
    \caption{Microhardness measurement}
    \label{fig:Microhardness}
\end{figure}

To assess the homogeneity of the swaged alloy, microhardness measurements were performed on transverse sections of the rods. After swaging to a diameter of \qty{9}{\milli\meter}, a pronounced hardness gradient was observed (Figure~\ref{fig:Microhardness}a). The hardness decreased from approximately 300--350 \qty{}{\HVhalf} at the rod surface to about \qty{250}{\HVhalf} at a depth of \qtyrange{1}{2}{\milli\meter}. This inhomogeneity may be associated with non-uniform induction heating. At the applied frequency of \qty{35}{\kilo\hertz}, the estimated electromagnetic skin depth is approximately \qtyrange{2.8}{3.1}{\milli\meter}, based on the temperature-dependent electrical resistivity of pure zirconium \cite{Desai1984ZrResistivity} and the conventional skin-depth relation \cite{Keller2023SkinEffect}. This suggests preferential electromagnetic heat generation near the rod surface, potentially resulting in a non-uniform temperature distribution during heating. Such thermal gradients may contribute to variations in the resulting microstructure and, consequently, to the observed hardness differences.

After subsequent swaging steps, the samples no longer exhibited any significant inhomogeneities in hardness. Their hardness remained essentially unchanged, with an average value of approximately \qty{280}{\HVhalf} and a standard deviation not exceeding \qty{10}{\HVhalf}.

\newpage

\section{Discussion} 
\subsection{Thermomechanical processing}
Thermomechanical processing with deformation temperatures selected relative to the $\beta$-transus is well established for $\alpha+\beta$ Ti alloys and has also been employed in Zr--Nb alloys to control phase constitution, microstructure, and texture \cite{Semiatin2020-wv,daniel_detailed_2019,Daniel2021-sq}. Rotary swaging has likewise been applied to Zr alloys \cite{Perlovich2008-pl,veverkova_novel_2021,Saibaba2013}, while induction heating has previously been coupled with rotary swaging of metallic billets \cite{Kocich2020-sw}. The present study combines these concepts into a multi-step hot rotary swaging route in which induction heating enables successive deformation within the single-phase $\beta$ and two-phase $\alpha+\beta$ fields. In addition, characterisation after the individual reductions provides direct information on the progressive development of microstructure and crystallographic texture throughout the processing sequence.

The experimentally determined transformation interval, extending from approximately \qty{550}{\degreeCelsius} to the $\beta$-transus, $T_{\beta}$, at approximately \qty{830}{\degreeCelsius}, was used to define the deformation temperatures. The $\beta$-field step was performed at $T_{\beta}+\qty{50}{\degreeCelsius}$, while subsequent swaging was conducted at $T_{\beta}-\qty{50}{\degreeCelsius}$. This approach follows the general principles established for sequential $\beta \rightarrow \alpha+\beta$ thermomechanical processing of Ti and Zr alloys, where deformation above $T_{\beta}$ is used to break down the coarse prior-$\beta$ structure and subsequent subtransus deformation promotes refinement and modification of the resulting $\alpha+\beta$ microstructure \cite{Semiatin2020-wv,Mallikarjun2003-ey,Lutjering1998-jf,Daniel2021-sq}.
\subsection{Microstructure evolution}
During swaging in the $\beta$ phase field, the initially millimetre-sized prior-$\beta$ grains were refined to approximately \qty{100}{\micro\meter}. Subsequent cooling produced a fine lamellar $\alpha+\beta$ microstructure. Both $\beta$-grain refinement during high-temperature deformation and the formation of lamellar $\alpha$ during cooling from the $\beta$ field are characteristic of $\beta$-processed Ti and Zr alloys \cite{Semiatin2020-wv,Fan2017-ij,Chakravartty2010-ok}.

Further swaging in the $\alpha+\beta$ field progressively broke down this transformation-induced lamellar structure. The $\alpha$ lamellae became increasingly globularised, while the retained $\beta$ grains underwent further refinement. The \qty{8}{\milli\meter} and \qty{7}{\milli\meter} states consisted predominantly of relatively equiaxed $\alpha$ grains separated by thin regions of $\beta$, resembling microstructures previously reported after hot forging and rolling of conventional Zr--Nb alloys \cite{Zhang2026-yl,Chai2018-dw,Daniel2021-sq}. These observations are consistent with the general tendency of deformation in the $\alpha+\beta$ field to fragment the lamellar structure and promote progressive globularisation of the $\alpha$ phase.

After the final swaging step to \qty{6}{\milli\meter}, the morphology changed further. The $\alpha$ grains became distinctly elongated along the swaging direction, whereas the $\beta$ phase developed a more globular morphology. Elongated $\alpha$ grains have also been observed after substantial hot deformation of Zr--1Nb and Zr--2.5Nb \cite{Zhang2026-yl,Rajaee2025-ow}, as well as during $\alpha+\beta$ rolling of model Zr--Nb alloys \cite{Daniel2021-sq}. However, the simultaneous development of a more globular $\beta$ morphology differs from the thin intergranular or elongated $\beta$ regions commonly reported after conventional $\alpha+\beta$ processing. The final state therefore indicates different morphological responses of the two constituent phases to the accumulated rotary-swaging deformation.

\subsection{Texture evolution}
The microstructural evolution was accompanied by pronounced changes in crystallographic texture. In the $\alpha$ phase, the basal poles progressively tilted away from the swaging direction. Such behaviour is characteristic of deformation textures in Zr and has previously been reported after rolling, drawing, and swaging \cite{Burgers1937_ZrWireTexture,Nerses1960_ZrTubeTexture,FU2023144584}. However, the $\langle10\bar{1}0\rangle_{\mathrm{hcp}}\parallel\mathrm{SD}$ component commonly associated with cold deformation was not observed. Instead, progressive swaging produced a $\langle11\bar{2}0\rangle_{\mathrm{hcp}}\parallel\mathrm{SD}$ fibre component. In Zr alloys, this orientation has been associated with recrystallisation textures developed after cold working and represents an approximately \qty{30}{\degree} rotation around the $c$-axis from the deformation texture \cite{Chaubet2002_EBSDRecrystallizationZircaloy4,Tenckhoff1970_AnnealingTexturesZircaloy,Cheadle1966_TextureZrRichAlloys}. Its development is therefore consistent with the microstructural evidence for recrystallisation during $\alpha+\beta$ swaging.

In the $\beta$ phase, a $\langle110\rangle_{\mathrm{bcc}}\parallel\mathrm{SD}$ component was present throughout the swaging sequence. This fibre texture is characteristic of axisymmetric deformation of bcc metals and has been reported after wire drawing and rotary swaging \cite{guo_microstructural_2013,naydenkin_structural-phase_2023}. Its development is consistent with crystallographic rotations associated with the common bcc $\{110\}\langle111\rangle$, $\{112\}\langle111\rangle$, and $\{123\}\langle111\rangle$ slip systems \cite{kocks_texture_2000}.

A $\langle111\rangle_{\mathrm{bcc}}\parallel\mathrm{SD}$ component was also observed, particularly after swaging to \qty{9}{\milli\meter} and \qty{6}{\milli\meter}. This component is less characteristic of a conventional bcc swaging texture and may be associated with recrystallisation of the $\beta$ phase. Preferential development of $\langle111\rangle$ components during recrystallisation has been reported in bcc alloys \cite{Zhang2016_Recrystallization9Cr1Mo}, while recrystallising $\beta$-Ti alloys subjected to forging or hot compression can likewise develop pronounced $\langle111\rangle_{\mathrm{bcc}}$ fibre components \cite{Rao2025_MDRXTi6242,Gu2021_Ti5321Texture}. Recrystallisation therefore clearly affects the observed $\beta$ texture evolution.

The recrystallisation-associated texture components observed in the two phases are also crystallographically related. The $\langle11\bar{2}0\rangle_{\mathrm{hcp}}\parallel\mathrm{SD}$ component in $\alpha$ and the $\langle111\rangle_{\mathrm{bcc}}\parallel\mathrm{SD}$ component in $\beta$ satisfy the Burgers orientation relationship, $\{0001\}_{\mathrm{hcp}}\parallel\{110\}_{\mathrm{bcc}}$ and $\langle11\bar{2}0\rangle_{\mathrm{hcp}}\parallel\langle111\rangle_{\mathrm{bcc}}$ \cite{Burgers1934}. The two components may therefore develop independently through recrystallisation or may be coupled through texture inheritance during the $\alpha\leftrightarrow\beta$ transformation. Texture inheritance and variant selection during these transformations have been demonstrated in Zr alloys \cite{Daymond2010_TextureInheritance,Romero2009_TextureMemory}. The present measurements do not allow these contributions to be separated, but the crystallographic correspondence indicates that deformation, recrystallisation, and phase transformation may all participate in establishing the final texture.

\subsection{Phase composition}
Despite the pronounced changes in morphology and texture, the $\beta$-phase weight fraction remained relatively stable during the swaging process, decreasing from approximately \qty{15}{\wtpercent} in the initial and \qty{9}{\milli\meter} states to \qty{12}{\wtpercent} in the final state. This is slightly higher than the approximately \qty{11}{\wtpercent} $\beta$ estimated for Zr--2.5Nb \cite{Muransky2013-tr}, but considerably below the approximately \qty{22}{\wtpercent}, \qty{41}{\wtpercent}, and \qty{68}{\wtpercent} estimated for Zr--5Nb, Zr--10Nb, and Zr--15Nb, respectively, from the experimentally determined volume fractions \cite{Cai2012ZrNb}. The present alloy therefore remained predominantly $\alpha$ with a moderate retained $\beta$ fraction throughout processing. No additional intermetallic phases were detected.

\subsection{Compression tests}
The compression tests show that the microstructural evolution during swaging is accompanied by a substantial change in mechanical response. Direct comparison with previous work is complicated by the limited number of studies combining rotary swaging of Zr alloys with subsequent room-temperature compression testing. Nevertheless, conventionally processed Zr--Nb alloys provide useful reference values. Cai et al.\ reported compressive yield strengths of approximately \qtyrange{345}{423}{\mega\pascal} for hot-worked Zr--2.5Nb, depending on loading direction \cite{Cai2009Compression}, while values of approximately \qty{358}{\mega\pascal}, \qty{483}{\mega\pascal}, and \qty{547}{\mega\pascal} were reported for Zr--5Nb, Zr--10Nb, and Zr--15Nb, respectively \cite{Cai2012ZrNb}. The present values of \qtyrange{839}{1103}{\mega\pascal} are therefore substantially higher. High strength has also been reported in other strongly processed Zr alloys: Singh et al.\ obtained a tensile yield strength of approximately \qty{674}{\mega\pascal} in rotary-swaged Zr--4 \cite{Singh2022-ce}, while Yang et al.\ reported tensile yield strengths of approximately \qtyrange{507}{558}{\mega\pascal} in severely deformed Zr--2.5Nb \cite{yang_achieving_2026}. These tensile data provide only contextual comparison because of the pronounced tension--compression asymmetry of textured Zr alloys \cite{Cai2009-ys,Holt2009-hb}.

Within the present processing sequence, the most pronounced change is the reduction in $\sigma_{0.2}$ from \qty{1103}{\mega\pascal} in the \qty{9}{\milli\meter} state to \qty{839}{\mega\pascal} in the \qty{7}{\milli\meter} and \qty{6}{\milli\meter} states. This evolution coincides with substantial changes in both texture and microstructure. The \qty{9}{\milli\meter} state contains a pronounced $\langle0001\rangle_{\mathrm{hcp}}\parallel\mathrm{DA}$ component, where the deformation axis (DA) during compression coincides with the swaging direction. Such an orientation is geometrically unfavourable for prismatic $\langle a\rangle$ slip, one of the deformation modes with the lowest critical resolved shear stress in $\alpha$-Zr \cite{Holt2009-hb}. Accommodation of strain parallel to the $c$ axis instead requires a greater contribution from mechanisms capable of producing a $c$-axis strain component, including pyramidal $\langle c+a\rangle$ slip and deformation twinning \cite{Cai2009-ys,Holt2009-hb}. Twinning during compression close to the $c$ axis has also been observed directly in Zr single crystals \cite{Akhtar1973-zt}. The strong $\langle0001\rangle_{\mathrm{hcp}}\parallel\mathrm{DA}$ component may therefore contribute to the high yield strength of the \qty{9}{\milli\meter} state.

Further swaging suppresses this component while increasing the contribution of $\langle11\bar20\rangle_{\mathrm{hcp}}\parallel\mathrm{DA}$ and $\langle10\bar11\rangle_{\mathrm{hcp}}\parallel\mathrm{DA}$ orientations. Previous studies of Zr--2.5Nb have demonstrated a pronounced dependence of yielding, slip, twinning, and interphase load partitioning on the orientation of the loading axis relative to the initial texture \cite{Cai2009-ys,Muransky2013-tr}. The observed texture evolution may therefore contribute to the progressive reduction in macroscopic yield strength.

\begin{figure}[ht]
\centering
\includegraphics[width=0.65\linewidth]{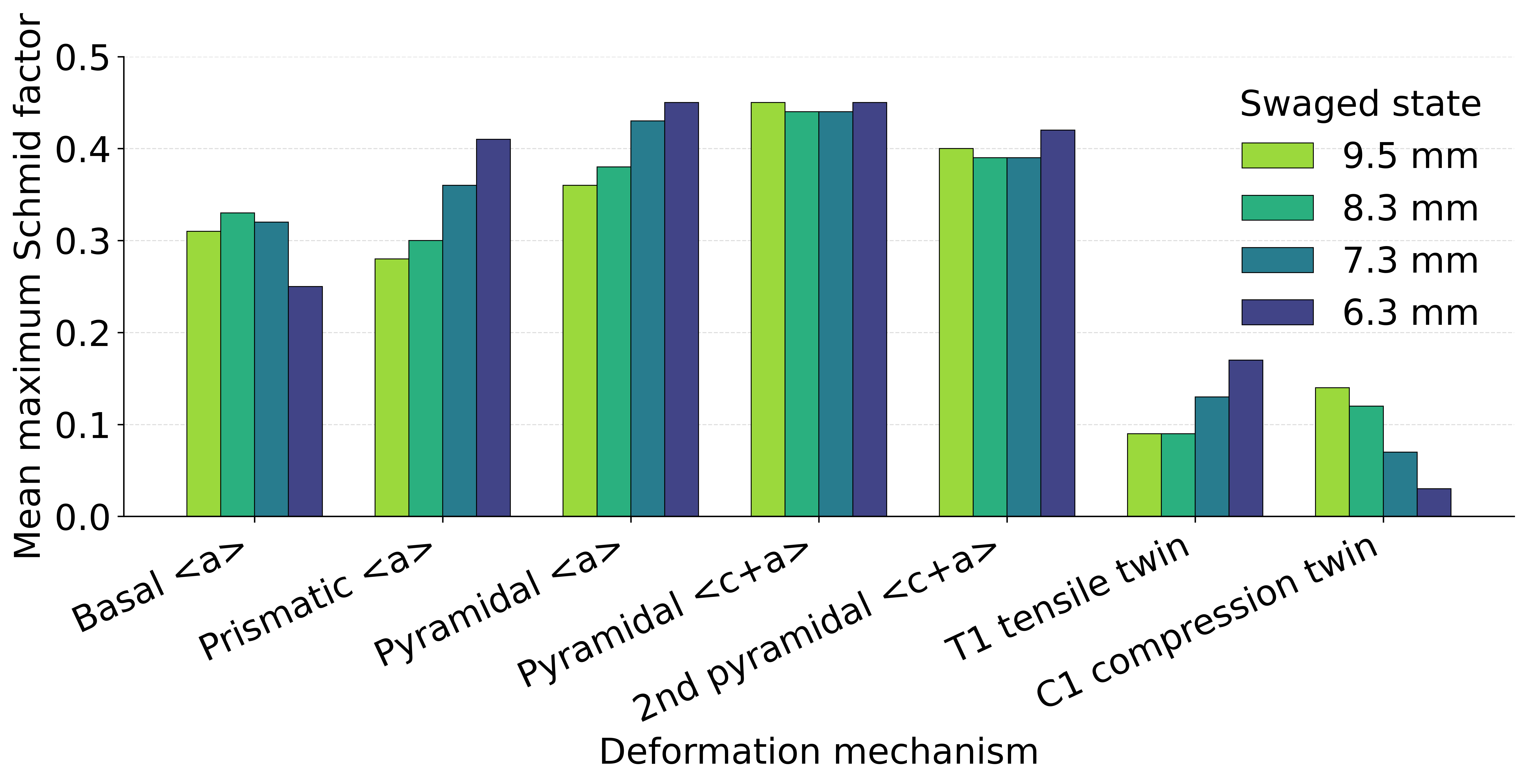}
\caption{Schmid factors for the selected hcp deformation mechanisms calculated from the ODF obtained by XRD texture measurements.}
\label{Schmidt factor}
\end{figure}

This interpretation is supported qualitatively by the Schmid-factor analysis shown in Figure~\ref{Schmidt factor}. The calculations were performed using 5000 orientations sampled according to the XRD-derived ODF, and for each deformation-mode family the maximum Schmid factor among the crystallographically equivalent variants was determined for each orientation and subsequently averaged. Progressive swaging increases the mean grain-wise maximum Schmid factors for prismatic $\langle a\rangle$ slip and for several of the considered pyramidal-slip and twinning systems. The texture thus provides increasingly favourable geometrical conditions for activation of these deformation modes.

The Schmid factors should, however, be interpreted only as geometrical indicators. They do not account for differences in critical resolved shear stress between deformation systems, grain-to-grain compatibility, interphase stresses, residual stresses, or the directional nature of deformation twinning \cite{Cai2009-ys,Holt2009-hb,Muransky2013-tr}. Consequently, their evolution supports a contribution of texture to the reduction in yield strength but does not identify the deformation mechanisms responsible for yielding.

Texture evolution is also unlikely to be the sole origin of the observed softening. During $\alpha+\beta$ swaging, the transformation-derived lamellar $\alpha$ colonies are progressively broken down, while globularisation and recrystallisation become increasingly pronounced. Similar behaviour has been observed during hot deformation of Zr--1Nb and Zr--2.5Nb, where dynamic recrystallisation and globularisation of initially lamellar structures occur within the $\alpha+\beta$ field \cite{Chakravartty1996-so,Sarkar2013-hv,Saxena2016-pb}. These restoration processes may reduce part of the deformation substructure retained from preceding swaging steps and thereby contribute to the decrease in yield strength $\sigma_{0.2}$. Their effect cannot, however, be separated from concurrent grain refinement and dislocation storage. Saxena et al.\ demonstrated that strength after two-phase deformation of Zr--1Nb depends simultaneously on $\alpha$ morphology, lath size, texture, and dislocation density \cite{Saxena2018-ho}. The decrease in yield strength in the present material should therefore be regarded as the combined result of texture evolution, restoration, and changes in grain and phase morphology.

The $\beta$ phase can additionally influence the macroscopic response through interphase load partitioning. In-situ diffraction studies of Zr--Nb alloys have shown progressive transfer of load from the softer $\alpha$ phase toward the harder $\beta$ phase during plastic deformation \cite{Cai2009-ys,Muransky2013-tr}. Mur\'{a}nsky et al.\ estimated a $\beta$-phase yield strength of approximately \qtyrange{750}{900}{\mega\pascal} in fine-grained Zr--2.5Nb \cite{Muransky2013-tr}, while Sahoo et al.\ demonstrated that changing the relative hardness of the $\beta$ phase substantially modifies lattice strain, twinning, and deformation development in the surrounding $\alpha$ phase \cite{Sahoo2010-pa}. Changes in $\beta$ morphology during the present processing sequence may therefore also alter the mechanical constraint imposed on the $\alpha$ matrix.

The evolution of the post-yield response shows that $\alpha+\beta$ swaging does not simply reduce the overall strength. While $\sigma_{0.2}$ decreases markedly, the UCS changes considerably less. The stress increment between yield and maximum compressive stress increases from approximately \qty{170}{\mega\pascal} in the \qty{9}{\milli\meter} state to \qty{280}{\mega\pascal} and \qty{360}{\mega\pascal} in the \qty{8}{\milli\meter} and \qty{7}{\milli\meter} states, respectively, while the fracture strain increases from \qty{3.6}{\percent} to \qty{32.2}{\percent}. The intermediate states therefore exhibit a substantially greater ability to accommodate plastic deformation while continuing to sustain increasing stress. Because the maximum stress is reached at different strains, the difference between UCS and $\sigma_{0.2}$ should not be interpreted as a quantitative strain-hardening parameter, but it clearly demonstrates the change in post-yield response.

The final swaging step produces a smaller but distinct change. Although $\sigma_{0.2}$ remains at \qty{839}{\mega\pascal}, the UCS decreases from \qty{1199}{\mega\pascal} to \qty{1131}{\mega\pascal} and the fracture strain decreases from \qty{32.2}{\percent} to \qty{26.4}{\percent}. The similar yield strengths of the \qty{7}{\milli\meter} and \qty{6}{\milli\meter} states suggest that the factors governing the onset of macroscopic plasticity remain similar, whereas their post-yield deformation differs. This change coincides with further texture evolution and with the transition from predominantly equiaxed $\alpha$ grains toward elongated $\alpha$ grains and a more globular $\beta$ morphology. The \qty{7}{\milli\meter} state therefore appears to provide the most favourable combination of high strength and plastic deformability within the investigated processing sequence.

Overall, the results demonstrate that the evolution during phase-field-controlled rotary swaging cannot be described by grain refinement alone. Sequential deformation in the $\beta$ and $\alpha+\beta$ fields produces coupled changes in phase morphology, recrystallisation state, crystallographic texture, and interphase constraint, which are reflected directly in the mechanical response. In particular, subsequent $\alpha+\beta$ swaging transforms the strong but poorly deformable \qty{9}{\milli\meter} condition into states with substantially improved plastic deformability while retaining high compressive strength.

\section{Conclusions}\label{sec:conclusions}

A multi-step hot rotary swaging route for the dual-phase Zr--4Nb--2Al alloy was successfully developed using induction heating and thermographic temperature control. Processing temperatures were selected based on the experimentally determined $\beta$-transus temperature, enabling sequential deformation in the single-phase $\beta$ field at \qty{880}{\degreeCelsius} and in the $\alpha+\beta$ field at \qty{780}{\degreeCelsius}. The evolution of the material was systematically investigated after the individual swaging steps up to a total area reduction of \qty{72}{\percent}.

Swaging in the $\beta$ field refined the initially millimetre-sized prior-$\beta$ grains and, after cooling, produced a fine lamellar $\alpha+\beta$ microstructure. Subsequent $\alpha+\beta$ swaging progressively globularised and recrystallised the $\alpha$ phase and further refined the $\beta$ phase. This was accompanied by pronounced texture evolution: the $\beta$ phase retained a characteristic $\langle110\rangle_{\mathrm{bcc}}\parallel\mathrm{SD}$ fibre component, while the $\alpha$ phase progressively developed a $\langle11\bar{2}0\rangle_{\mathrm{hcp}}\parallel\mathrm{SD}$ texture. The room-temperature phase constitution remained stable, with approximately \qtyrange{12}{15}{\wtpercent} retained $\beta$ and no additional phases detected.

The processing stages also produced substantial changes in mechanical behaviour. The compressive yield strength decreased from \qty{1103}{\mega\pascal} to \qty{839}{\mega\pascal} during subsequent $\alpha+\beta$ swaging, while the ability to accommodate plastic deformation increased markedly, with the maximum fracture strain of \qty{32.2}{\percent} reached in the \qty{7}{\milli\meter} state. At the same time, the initially pronounced radial hardness heterogeneity was progressively eliminated, resulting in an approximately homogeneous final hardness of \qty{280}{\HVhalf}.

Overall, the results demonstrate the feasibility of induction-assisted, multi-step hot rotary swaging for processing dual-phase Zr alloys and establish how their microstructure, texture and mechanical response evolve throughout the individual deformation stages. The resulting combination of high compressive strength and substantial plastic deformability highlights the potential of this processing approach for developing dual-phase zirconium alloys intended for nuclear applications.

\section*{Acknowledgements} 
This work was supported by the Czech Science Foundation (GAČR) under projects No. 26-21789S, and by the Grant Agency of Charles University (GAUK) under project No. 272523. Partial financial support was provided by the Ministry of Education, Youth and Sports of the Czech Republic through project No. CZ.02.01.01/00/22\_008/0004591, co-funded by the European Union.

Part of the experimental work was carried out at the Materials Growth and Measurement Laboratory (MGML, \href{https://mgml.eu}{mgml.eu}), supported by the Czech Research Infrastructures programme (project No. LM2023065).

\section*{Data availability}
The data that support the findings of this study are openly available in the Zenodo repository at \href{https://doi.org/10.5281/zenodo.23345748}{https://doi.org/10.5281/zenodo.23345748}  under the CC-BY 4.0 license.

\section*{Declaration of generative AI and AI-assisted technologies in the manuscript preparation process.}
During the preparation of this work, the authors used ChatGPT (OpenAI) to improve the language and readability of the manuscript, assist in the identification of relevant scientific literature, and support the development and debugging of Python and MATLAB scripts used for experimental data processing. All identified literature was independently read and verified by the authors. After using this tool, the authors reviewed and edited the content as needed and take full responsibility for the content of the published article.

\input AAArticle.bbl

\newpage
\section*{Supplementary Material}

\setcounter{figure}{0}
\setcounter{table}{0}

\renewcommand{\thefigure}{S\arabic{figure}}
\renewcommand{\thetable}{S\arabic{table}}

\begin{figure}[h!]
    \centering
    \includegraphics[width=0.95\linewidth]{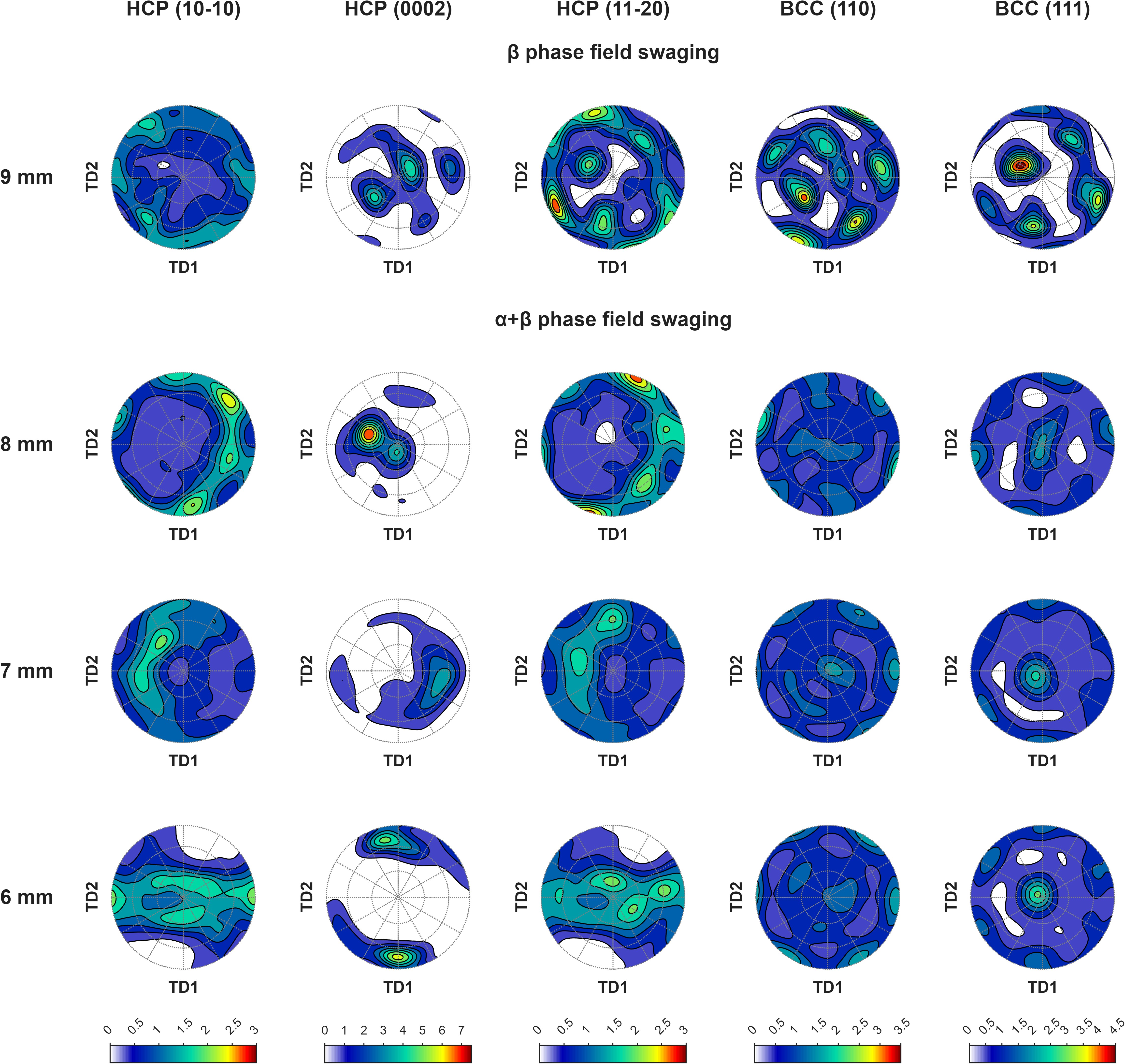}
    \caption{Pole figures obtained from EBSD measurements. The pole figures are displayed in the plane of transverse directions TD1--TD2, with the swaging direction (SD) corresponding to the centre of each pole figure.}
    \label{EBSD PF grid}
\end{figure}

\begin{figure}[h!]
    \centering
    \includegraphics[width=0.95\linewidth]{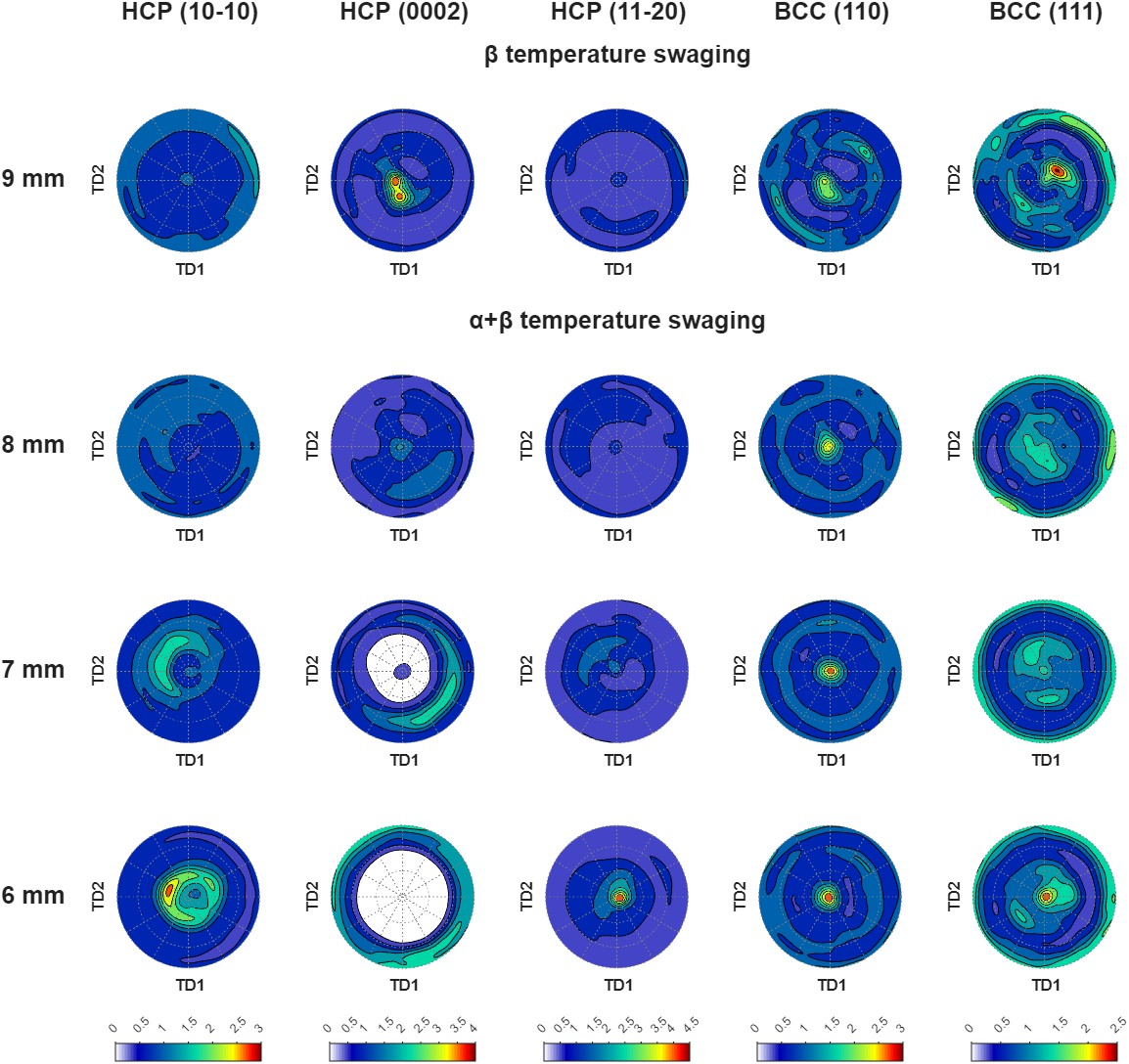}
    \caption{Calculated PFs from XRD texture measurement using Schulz method.}
    \label{XRD texture}
\end{figure}

\begin{table}[h!]
\centering
\begin{tabular}{|c|c|c|c|c|c|c|c|c|} \hline
State
& $R$-NP (\%)
& $R_{\mathrm{wp}}$-NP (\%)
& GOF-NP
& DW-NP
& $R$-FP (\%)
& $R_{\mathrm{wp}}$-FP (\%)
& GOF-FP
& DW-FP \\ \hline

Initial
& 9.64 & 12.78 & 4.36 & 0.288
& 9.64 & 12.78 & 4.36 & 0.288 \\

9 mm
& 9.62 & 12.12 & 5.73 & 0.167
& 9.62 & 12.13 & 5.73 & 0.167 \\

8 mm
& 9.30 & 11.94 & 5.56 & 0.203
& 9.28 & 11.94 & 5.56 & 0.202 \\

7 mm
& 9.31 & 11.67 & 5.20 & 0.243
& 9.31 & 11.68 & 5.20 & 0.242 \\

6 mm
& 9.07 & 11.44 & 4.60 & 0.239
& 9.07 & 11.44 & 4.61 & 0.239 \\ \hline
\end{tabular}
\caption{Quality-of-fit parameters obtained for the no-partition (NP) and full-partition (FP) Rietveld refinements. DW denotes the Durbin--Watson statistic.}
\label{tab:rietveld_fit_quality}
\end{table}

\end{document}